%% file: main.tex
\documentclass{article}
\usepackage[utf8]{inputenc}
\usepackage[english]{babel}
\usepackage{graphicx}
\usepackage{amsmath}
\usepackage{amsthm}
\usepackage{amsfonts}
\usepackage{amsopn}
\usepackage{amssymb}
\usepackage{mathtools}
\usepackage{xcolor}
\usepackage{listings}
\usepackage{appendix}
\usepackage{booktabs}
\usepackage[symbol]{footmisc}

\newtheorem{definition}{Definition}[section]
\newtheorem{example}{Example}[section]

\usepackage{algorithm}
\usepackage{algorithmic}

\usepackage[nottoc,numbib]{tocbibind}
\usepackage[hidelinks]{hyperref}

\usepackage[a4paper,top=3cm,bottom=2cm,left=3cm,right=3cm,marginparwidth=1.75cm]{geometry}

\input{macros.tex}

\title{Accelerated Sampling on Discrete Spaces with Non-Reversible Markov Processes}
\author{Sam Power\footnote{Department of Pure Mathematics and Mathematical Statistics, University of Cambridge. SP acknowledges financial support from the Cantab Capital Institute for the Mathematics of Information.}, Jacob Vorstrup Goldman\footnote{Signal Processing and Communications Group, Department of Engineering, University of Cambridge. JVG acknowledges financial support from an EPSRC Doctoral Training Award.}}

\begin{document}

\maketitle

\begin{abstract}
\begin{itemize}
We consider the task of MCMC sampling from a distribution defined on a discrete space. Building on recent insights provided in \cite{zanella2019informed}, we devise a class of efficient continuous-time, non-reversible algorithms which make active use of the structure of the underlying space. Particular emphasis is placed on how symmetries and other group-theoretic notions can be used to improve exploration of the space. We test our algorithms on a range of examples from statistics, computational physics, machine learning, and cryptography, which show improvement on alternative algorithms. We provide practical recommendations on how to design and implement these algorithms, and close with remarks on the outlook for both discrete sampling and continuous-time Monte Carlo more broadly.
\end{itemize}
\end{abstract}

\tableofcontents
\addtocontents{toc}{~\hfill\textbf{Page}\par}

\section{Introduction}
\input{Inputs/A_Introduction/general_intro.tex}

\section{Continuous-time Algorithms on Discrete Spaces}
\subsection{Motivation: Locally-Balanced MCMC in Continuous Time}
\input{Inputs/B_Algorithmic_Contributions/cZanella.tex}

\subsection{Simulation on Spaces with Algebraic Structure}
\input{Inputs/B_Algorithmic_Contributions/Groups.tex}
\subsubsection{Designing Non-Reversible Markov Processes on Discrete Spaces}
\input{Inputs/B_Algorithmic_Contributions/GroupsSkewReversible.tex}

\subsection{Tabu Sampler: Self-Avoiding Walks on Spaces With Low Order Generators}
\input{Inputs/B_Algorithmic_Contributions/SAW.tex}

\subsection{Persistent Piecewise Deterministic Markov Processes for Spaces With High Order Generators}
\input{Inputs/B_Algorithmic_Contributions/PDMP.tex}

\section{Numerical Examples}
In this section we first discuss implementation choices for practitioners, and subsequently provide numerical evidence of the performance of the Tabu, dC and dZZ samplers. Code in Python 3 for all examples are available online at \href{https://github.com/jvorstrupgoldman/tabu_dc_dzz}{GitHub}.
\subsection{Implementation of Continuous-time Samplers}
\input{Inputs/D_Numerical_Examples/Implementation.tex}
\subsubsection{Choice of Balancing Function}
\input{Inputs/D_Numerical_Examples/LocallyBalancing.tex}
%

\subsection{Examples with Low Order Generators}\label{sec:low_order}
\input{Inputs/D_Numerical_Examples/LowOrder.tex}
\subsubsection{Bayesian Variable Selection}\label{section:bayvar}
\input{Inputs/D_Numerical_Examples/BayVar.tex}
\subsubsection{Conditional Permutation Test}\label{seq:cpt}
\input{Inputs/D_Numerical_Examples/CPerm.tex}
\subsubsection{Spin Glasses: The Sherrington-Kirkpatrick Model}
\input{Inputs/D_Numerical_Examples/SpinGlass.tex}
\subsubsection{Log-Submodular Distributions I: Facility Location}
\input{Inputs/D_Numerical_Examples/SubPoint.tex}
\subsubsection{Log-Submodular Distributions II: Determinantal Point Processes}\label{seq:dpp}
\input{Inputs/D_Numerical_Examples/DetPoint.tex}
\subsection{Examples with High Order Generators}
\input{Inputs/D_Numerical_Examples/HighOrder.tex}
\subsubsection{Lattice Gaussians}
\input{Inputs/D_Numerical_Examples/LatticeGaussian.tex}
\subsubsection{Discrete Lattice Gauge Theories}
\input{Inputs/D_Numerical_Examples/LatticeGauge.tex}
\section{Conclusion and Outlook}
\input{outlook.tex}
\clearpage

\bibliographystyle{alpha}
\bibliography{bibliography}

\clearpage
\appendix
\section{Proofs of Invariance}
\input{Inputs/E_Appendix_Bibliography/InvarianceProofs.tex}

\end{document}

%% file: macros.tex
\newcommand{\Lc}{\mathcal{L}}

\newcommand{\Qc}{\mathcal{Q}}

\newcommand{\Uc}{\mathcal{U}}

\newcommand{\Xc}{\mathcal{X}}

\newcommand{\Eb}{\mathbf{E}}

\newcommand{\Pb}{\mathbf{P}}

\DeclareMathOperator*{\argmax}{arg\,max}

%% file: Inputs/A_Introduction/general_intro.tex
In this work, we concern ourselves with Markov Chain Monte Carlo (MCMC) simulation of distributions defined on discrete state spaces. 

On continuous spaces, there has been a great deal of successful work on how to construct efficient MCMC proposals which work in great generality. Much of this success has stemmed from identifying continuous-time dynamical processes (ODEs, SDEs, PDMPs) which admit the desired invariant measure, and then discretising those processes to form tractable discrete-time chains.

This approach has apparently seen less use in the discrete setting. A reasonable justification for this is that differential equations do not exist per se on discrete spaces, and so extending this approach to derive appropriate dynamics for discrete spaces is perhaps unclear at first. However, one can note the following: it is essentially the case that, on discrete spaces, the various notions of Markov process on continuous spaces (ODEs, SDEs, PDMPs, etc.) all collapse to the same notion, that of a \textit{Markov Jump Process} (MJP). 

From the perspective of algorithm design, this presents a great simplification: any continuous time sampler on a discrete space must arise as an MJP. The first contribution of this work is to explicitly construct a family of MJPs which admit a desired invariant measure. Moreover, in contrast to the usual scenario on continuous state spaces, it is possibly to dispense entirely with discretisation and to simulate the process exactly in continuous time. In addition to motivating this general-purpose sampling algorithm for discrete spaces, which we term the \textit{Zanella Process}, this construction will allow us to retroactively justify the success of certain existing approaches to discrete sampling.

A second aspect of discrete sampling which is worthy of attention is how one can improve the computational efficiency of a sampler by making use of the symmetric and algebraic structure available in a state space. In particular, it is often the case that a state space $\mathcal{X}$ is naturally acted upon by a \textit{group} of symmetries $G$. Given such structures, which abound in applications, the natural question is how to best exploit them algorithmically. The second contribution of this work is to present a trio of algorithms which offer a generic solution to this question. The first, which we term the \textit{Tabu Sampler} is adapted to settings where $G$ is generated by low-order elements, and encourages the process to avoid re-using generators across short-to-medium timescales, thus preventing backtracking behaviour. 

In contrast, the latter two algorithms focus on the scenario where the group is generated by elements of high or infinite order, and instead encourage persistent motion across the state space, re-using generators so long as they are driving the dynamics in productive directions. The two schemes presented are the \textit{discrete Coordinate Sampler} (dCS) and \textit{discrete Zig-Zag Process} (dZZ), each named by analogy with corresponding algorithms in the literature on Piecewise-Deterministic Markov Processes (PDMPs), namely the works of \cite{wu2018coordinate, bierkens2019zig}.

All three of the algorithms are devised to be non-reversible, and as been witnessed in a range of applications, this appears to have a beneficial effect upon convergence behaviour.

We demonstrate that all of the schemes above are asymptotically exact in the standard MCMC sense, in each case by establishing either reversibility or skew-reversibility. We explain how to implement these schemes in practice, and provide a number of numerical experiments, as well as code. We also offer concrete recommendations on how and when each of these samplers should be applied in practice. We close with some discussion of potential future work in this area.

%% file: Inputs/B_Algorithmic_Contributions/cZanella.tex
In this section, we construct the \textit{Zanella process} algorithm for sampling from a distribution $\pi$ supported on a discrete space $\mathcal X$. We begin by introducing the notion of a \textit{Markov Jump Process} (MJP).

\begin{definition}
A Markov Jump Process is a continuous-time Markov process taking values in a countable state space $\Xc$. Such a process is characterised by its jump rates, $\lambda ( x \to y )$, defined such that
\begin{align}
    \Pb ( X_{t + h} = y | X_t = x ) &= h \cdot \lambda ( x \to y ) + o ( h ) \quad \text{for } y \neq x \\
    \Pb ( X_{t + h} = x | X_t = x ) &= 1 - h \cdot \Lambda ( x ) + o ( h )
\end{align}
where $\Lambda ( x ) = \sum_y \lambda ( x \to y )$. The \textit{generator} of such a Markov process is given by
\begin{align}
    \Lc f ( x ) = \sum_y \lambda ( x \to y ) \left[ f ( y ) - f ( x ) \right].
\end{align}
\end{definition}

We will also require the notion of \textit{reversibility} as it pertains to MJPs.

\begin{definition}
A Markov Jump Process with generator $\Lc$ is reversible with respect to the measure $\pi$ if, for all $f, g \in L^2 ( \pi )$, it holds that
\begin{align}
    \Eb_\pi \left[ ( \Lc f ) ( x ) g ( x ) \right] = \Eb_\pi \left[ f ( x ) ( \Lc g ) ( x ) \right].
\end{align}
By considering $f ( x ) = \mathbf{I} [ x = a ], g ( x ) = \mathbf{I} [ x = b ]$, one can show that this is equivalent to
\begin{align}
    \pi ( x ) \lambda ( x \to y ) = \pi ( y ) \lambda ( y \to x ) \quad \text{for all } x, y \in \Xc
\end{align}
\end{definition}

The utility of reversibility is that it is a sufficient condition for the MJP to be reversible with respect to $\pi$, and moreover, it is a local condition, and is hence easily verifiable. As such, a standard approach to constructing Markov processes with a given invariant measure is to explicitly construct a process which is reversible with respect to that measure. We now proceed along these lines.

In order to make our search for such a process more tractable, we make two simplifications. Firstly, we assume that our discrete space $\mathcal{X}$ admits the structure of a \textit{graph}, that is, there is some set of \textit{edges} $\mathcal{E} \subset \mathcal{X} \times \mathcal{X}$ which encodes the notion of locality in the space. If $(x, y) \in \mathcal{E}$, we will call $x$ and $y$ \textit{neighbours}, and interpret these points as being `close', in some suitable sense. We use the shorthand $\partial x = \{ y \in \mathcal{X} : (x, y) \in \mathcal{E} \}$ to denote the \textit{neighbourhood} of $x$, i.e. the set of all points in $\mathcal{X}$ which are adjacent to $x$.

Secondly, we will restrict ourselves to considering Markov Jump Processes (MJPs) $(X_t)_{t \geqslant 0}$ on $\mathcal X$ which satisfy the following desiderata:
\begin{enumerate}
    \item The process can only jump from $x$ to $y$ if $y \in \partial x$.
    \item For $x \in \mathcal X, y \in \partial x$, the jump rate from $x$ to $y$, $\lambda (x \to y)$, is purely a function of $\frac{ \pi (y) }{ \pi (x) }$.
    \item $X_t$ is in detailed balance with respect to $\pi$.
\end{enumerate}
The first of these ensures that $X_t$ respects the graphical structure of the model. The second is a simplifying assumption which streamlines the analysis, and can be weakened\footnote{For greatest generality, one could take $\lambda (x \to y)$ to be the product of a function of $\frac{ \pi (y) }{ \pi (x) }$ with a symmetric function $S ( x, y ) $.}. The third is sufficient to ensure that $X_t$ be ergodic with respect to $\pi$ (under mild additional assumptions), and is crucial for the validity of the procedure as a Monte Carlo method.

As established earlier, the third condition can be written as
\begin{align}
    \forall (x, y) \in E, \quad \pi (x) \lambda (x \to y) = \pi (y) \lambda (y \to x).
\end{align}
Applying the second condition, we write $\lambda (x \to y) = g \left( \frac{ \pi (y) }{ \pi (x) } \right)$ for some function $g$, and thus see that
\begin{align}
    \pi (x) g \left( \frac{ \pi (y) }{ \pi (x) } \right) &= \pi (y) \cdot g \left( \frac{ \pi (x) }{ \pi (y) } \right) \\
    g \left( \frac{ \pi (y) }{ \pi (x) } \right) &= \frac{ \pi (y) }{ \pi (x) } \cdot g \left( \frac{ \pi (x) }{ \pi (y) } \right) \\
    g ( t ) &= t \cdot g ( 1 /t ) \quad \text{for } t = \frac{ \pi (y) }{ \pi (x) }.
\end{align}
As such, we deduce that for an algorithm of this form to satisfy all of our desiderata, irrespective of the target distribution $\pi$, it is necessary and sufficient that $\lambda (x \to y) = g \left( \frac{ \pi (y) }{ \pi (x) } \right)$ for some function $g$ satisfying $g ( t ) = t \cdot g ( 1 /t )$. We refer to such $g$ as \textit{balancing functions}. As such, for the remainder of this section, we fix a balancing function $g$, and study the resulting process, which we term the \textit{Zanella process}, by analogy with the work of \cite{zanella2019informed}.
\begin{algorithm}\caption{Zanella Process for sampling from $\pi ( x ), x \in \mathcal{X} $ }\label{alg:cz}
\begin{enumerate}
\item At $x \in \mathcal{X}$,
    \begin{enumerate}
        \item For $y \in \partial x$, compute $\lambda ( x \to y ) = g \left( \frac{ \pi (y) }{ \pi (x) } \right)$.
        \item Compute $\Lambda ( x ) = \sum_{y \in \partial x} \lambda ( x \to y )$.
        \item Sample a waiting time $T \sim \text{Exponential} ( \text{rate } = \Lambda ( x) )$.
        \item Sample a new location $y \in \partial x$ with probability $\frac{ \lambda ( x \to y ) }{ \Lambda ( x ) }$.
        \item Advance time by $T$.
        \item Jump to $y$.
    \end{enumerate}
\end{enumerate}
\end{algorithm}

Informally, the process can be seen as a random walk which observes its neighbours, weighs up which of the neighbours is most preferable as a next location, and then chooses to jump there after a random amount of time. Note that by operating directly in continuous time, the algorithm is `rejection-free'. 

A reasonable analogy in continuous state spaces is the (overdamped) Langevin process, in the sense that both processes are `weakly greedy'; they naturally gravitate towards regions of higher probability, while retaining the ability to explore regions of lower probability from time to time. This analogy is developed further in the work \cite{livingstone2019robustness}.

The optimal choice of $g$ is not necessarily clear a priori. Empirically, some sensible choices include $g (t) \in \left\{ \sqrt{t}, \min (1, t), \frac{t}{1+t} \right\}$. We note that in \cite{zanella2019informed}, there seems not to be a universally-optimal choice of $g$; it appears to be genuinely task-dependent. We are unsure of whether the same reasoning holds for the continuous-time process. Moreover, as is often the case with continuous-time processes, if one wants to compare different choices of $g$, it is necessary choose a normalisation of some sort. This is because for any balancing function $g$, one could equally take $2g$ as the balancing function, and obtain a process which converges twice as fast - by the clock of the process, but certainly not in real life! Upon selecting a normalisation (e.g. $g (1) = 1$), it may be possible to derive an optimality result, in the spirit of \cite{peskun1973optimum}.

The chief cost of this algorithm is the repeated computation of expressions of the form $\frac{ \pi (y)}{ \pi (x) }$. Many target distributions of interest admit convenient factorisation structures, and in these settings, computing $\frac{ \pi (y)}{ \pi (x) }$ can be considerably cheaper than computing either of $\pi (x), \pi (y)$ (even up to a normalising constant). As such, when wall-clock time is a concern, it is imperative to carry out these computations judiciously. In situations where one does not have such simplifications, or where the chosen graphical structure is dense (i.e. each state has many direct neighbours), then the cost can grow somewhat. We expect that it is still generally worthwhile to work with the Zanella Process (as opposed to Random Walk-based algorithms), but it is difficult to say anything concrete at this level of generality. We present the Zanella process as a worthy baseline for general discrete sampling tasks, relative to random-walk based samplers. We have found it to be simple, transparent, and reliable on unimodal tasks, in particular for Bayesian sampling with highly-informative posterior distributions.

In terms of related work, the Zanella process bears a strong connection to \textit{Kinetic Monte Carlo} algorithms (also `KMC', see e.g. \cite{e2019applied}) which have long been used in the computational physics community, but which may not be well-known to statisticians. An early example of this is the `$N$-Fold Way' of \cite{bortz1975new}, which can be viewed as a discrete-time analog of the Zanella process. The `Waiting Time Method' (WTM) of \cite{dall2001faster}, was derived by embedding the $N$-Fold Way into continuous time, and is almost directly identical to the Zanella process. We learned of these references through the recent work of \cite{baldassi2017method}, which essentially puts the WTM approach back into discrete time, but with some useful algorithmic simplifications. Our use of balancing functions was motivated by the work of \cite{zanella2019informed}, which also prompted our interest in structured discrete sampling more broadly.

%% file: Inputs/B_Algorithmic_Contributions/Groups.tex
In the design of the Zanella process above, the implicit assumption of a graphical structure of the space $\mathcal X$ is made. It is often clear what the graphical structure is, but there is no guarantee that it facilitates sampling some arbitrary distribution $\pi$ on $\mathcal X$. Often the form of $\pi$ implies that a subset of edges are much more natural to travel along than others, and the graph has no inherent notion of how to highlight these directions. We will in this section see that, in many applications, the state space $\mathcal{X}$ is, in addition, also naturally furnished with an algebraic structure. That is, there exist a group $G$ which acts on $\mathcal{X}$. Informally, this means that there is a set $G$ of `actions' $g$ such that for any $x \in \mathcal{X}$, we can make sense of what it means to perform the action $g$ to the state $x$ and obtain a new state $y = $ `$g\star x$', where we by the latter mean applying the action $g$ to $x$. To make this more rigorous, we first recall the definition of a group.
\begin{definition}
A group is a set $G$, equipped with a binary operation $\bullet : G \times G \to G$, satisfying the four group axioms:
\begin{enumerate}
    \item For all $g, h \in G$, $g \bullet h \in G$. (Closure)
    \item For all $g, h, k \in G$, $( g \bullet h ) \bullet k = g \bullet ( h \bullet k )$. (Associativity)
    \item There is an element, the identity element, $id \in G$ such that for all $g \in G$, $g \bullet id = id \bullet g = g$.
    \item For all $g \in G$, there is an element $h = g^{-1} \in G$ such that $ g \bullet h = h \bullet g = id$.
\end{enumerate}
We denote a group by $(G, \bullet)$.
\end{definition}

In what follows, we will generally write the composition of group elements as $g \cdot h$ rather than $g \bullet h$ to reduce clutter, in practice overloading the multiplication operator. 

To motivate our approach, we will consider a simple but relevant case which illustrates how a space can be associated with a group. Begin by considering a set $\mathcal S$ of $n$ elements, i.e., a set isomorphic to $[n] \equiv \{1,2, \cdots, n-1, n\}$ and consider the set of all one-to-one functions from $\mathcal S$ to itself. If we equip this set with the operation of function composition, this is known as the symmetric group $\Sigma_{ n} \equiv (\Sigma_n, \circ)$. The elements in $\Sigma_{n}$ corresponds to all permutations of the objects in $\mathcal S$, so the symmetry group gives us a natural way to travel everywhere between points of $\mathcal S$ by selecting the components of $\Sigma_{ n}$ appropriately. 

Nonetheless, from an algorithmic perspective, the symmetry group is often vast and unmanageable, as the order (number of elements) of $\Sigma_{ n}$ is $n!$. It is therefore much more useful to identify, if possible, a subset consisting of the `fundamental' elements, i.e. the building blocks of the group, allowing us to decompose any permutation into a series of simple steps. This is akin to the notion of a basis for a vector space, and such a collection is known as a \textit{generating set}:
\begin{definition}
Let $(G, \cdot)$ be a group, and let $\Gamma \subset G$ be a subset. Define $\langle \Gamma \rangle$ recursively by
\begin{enumerate}
    \item $\Gamma \subset \langle \Gamma \rangle$.
    \item $\gamma \in \langle \Gamma \rangle \implies \gamma^{-1} \in \langle \Gamma \rangle$.
    \item $\gamma, \eta \in \langle \Gamma \rangle \implies \gamma \cdot \eta, \eta \cdot \gamma \in \langle \Gamma \rangle$.
\end{enumerate}
Then $(\langle \Gamma \rangle,  \cdot)$ is a subgroup of $(G, \cdot)$. If $\langle \Gamma \rangle = G$, we say that $\Gamma$ is a generating set for $G$ and call the elements of $\Gamma$ generators. A generating set $\Gamma$ is furthermore \textsc{symmetric} if $\gamma \in \Gamma \implies \gamma^{-1} \in \Gamma$. 
\end{definition}
Thus with a generator set $\Gamma$ it is possible to decompose any element $g \in G$ to a series of simple group operations. For sampling purposes we now need, given a generator set of a particular group, a way for the generators to be sequentially applied to states $x\in \mathcal X$, rather than just applied to generate complicated elements of $G$. The concept of a \textit{group action} makes this heuristic formal:
\begin{definition}
Given a set $\mathcal{X}$ and a group $(G, \cdot)$, a \textit{group action} is defined as an operation $\star$ mapping $G \times \mathcal{X}$ to $\mathcal{X}$ satisfying:
    \begin{enumerate}
        \item $\forall x \in \mathcal{X}, \quad \text{id} \star x = x$.
        \item $\forall g, h \in G, \quad g \star ( h \star x )  = (g \cdot h) \star x $.
    \end{enumerate}
\end{definition}
From the definition, we can see that group actions are a natural way in which to introduce movement between states of a space $\mathcal X$. The samplers we introduce below will in every case carry out exploration on the target space via repeated applications of group actions to $\mathcal X$. Since objects in $\Gamma$ generate $G$, we can furthermore restrict ourselves to only consider group actions of the generators. However, the form of the generating set can often be determined by the specific application we consider rather than generically being the same for the same latent space. In order to make these notions somewhat more concrete, we first provide two examples equivalent to the space $\mathcal X = [n]$ we considered above, but where the target distribution of interest induce significantly different generator sets:

\begin{example}[Matching/Linkage (e.g. \cite{berrett2018conditional, zanella2019informed})] \label{example:linkage}
Suppose we have two sets of covariates, $\{ Y_i \}_{i = 1}^N$ and $\{ Z_j \}_{j = 1}^N$, which correspond to the same individuals, but in an unknown order. A natural task is to align these two sets, i.e. to identify a bijective mapping $\sigma : [n] \to [n]$ such that $Y_i$ and $Z_{\sigma ( i ) }$ correspond to the same individual. This set of maps is just the symmetric group on $n$ elements $\Sigma_n$, introduced above. As $\mathcal{X}$ is already a group in this case, the natural choice is to take $G$ to be the same group, i.e. to set $G = \Sigma_n$. Since there is no ordering structure to the covariates, a reasonable choice of generating set $\Gamma$ is the set of all transpositions, i.e. mappings $\tau_{i, j}$ which results in group actions given by
%
%
%
\begin{align}
    y = \tau_{i, j} \star x \Longleftrightarrow
        \begin{cases}
            y_j = x_i \\ 
            y_i = x_j \\
            y_k = x_k \quad \text{ for $k \neq i, j$}.
        \end{cases}
\end{align}
In words, applying the action $\tau_{i, j}$ to $x$ would thus correspond to swapping our beliefs about which $Z$ covariates correspond to $Y_i$ and $Y_j$ respectively. The cardinality of $|\Gamma|$ in this case is $n(n-1)$.
\end{example}

\begin{example}[Ranking]
Suppose now that we have $n$ individuals, and we want to rank them in some way. As in the matching case, the underlying state space is $\mathcal{X} = \Sigma_n$, and it is natural to take $G = \Sigma_n$ as well. However, because the ordering of $[n]$ is now meaningful for our problem, it is more natural to use the set of all \textsc{adjacent} transpositions as a generating set for our task, i.e. to use $\{ \tau_{i, j} \}_{|i - j| = 1}$. Applying the mapping $\tau_{i, i +1 }$ then corresponds to swapping our beliefs about the relative rankings of the individuals who were previously ranked in $i^{\text{th}}$ and $(i + 1)^{\text{th}}$ position. This generator set is just of cardinality $(n-1)$. 
\end{example}

The above examples illustrates how different algebraic structures can be exploited to serve our goal of sampling a particular distribution, and also be used to make the set of possible directions much more manageable in size. We now consider other spaces where the group is significantly different from the symmetry group.

\begin{example}[Binary Spin Systems]\label{spinglass_action}
In statistical physics, one commonly studies binary spin systems on graphs. Here, there is an underlying graph $G_0 = (V, E)$, and for each vertex $i$, there is a binary `spin', $\sigma_i \in  \{ \pm 1 \}$. The state space of interest is then given by $\mathcal{X} = \{ \pm 1 \}^{|V|} $. The simplest group action one can define for such a system is generated by picking a vertex $i$, and flipping the spin at that vertex, i.e. setting $\sigma'_i = - \sigma_i$. If we call that flipping move $\gamma_i$, then $\Gamma = \{ \gamma_i \}_{i \in V}$ with cardinality $|V|$ generates a group $(G, \cdot)$ which acts on $\mathcal{X}$. This group is isomorphic to the product of cyclic groups, $\mathbb{Z}_2^{|V|}$, and is often directly identified as such.
\end{example}

\begin{example}[Subset Selection]
Fix a finite set $S$, and consider the task of selecting a subset $A \subset S$. The state space is then given by the power set of $S$, which is naturally isomorphic to $\{ 0, 1 \}^S$, by taking a $1$ in the $i^{\text{th}}$ coordinate to mean `element $i$ is included in the subset $A$'. Mathematically, this is now essentially equivalent to the spin system setting; one can define $\gamma_i$ to act as
\begin{align}
    \gamma_i \star A = 
    \begin{cases}
    A \backslash\{ i \} \quad \text{ if } i \in A \\
    A \cup \{ i \} \quad \text{ if } i \notin A,
    \end{cases}
\end{align}
i.e. delete $i$ if it is already in $A$, otherwise include it. Again, taking $\Gamma = \{ \gamma_i \}_{i \in S}$ generates a group $(G, \cdot)$ which acts on $\mathcal{X}$ and is isomorphic to $\mathbb{Z}_2^V$.
\end{example}

\begin{example}[Lattice Distributions]\label{ex:high_order}
Consider a lattice $L \subseteq \mathbb{Z}^n$ for some integer $n$, and a distribution $\pi$ defined over all points of $\mathcal X = L$. Most of the classical discrete distributions used in probability and statistics are of this form. In this case, for a given point $x \in L$, we can define a move as either increasing or decreasing the value of a particular index $x_i$, $i=1,2,\ldots, n$:
\begin{align}
    y = \gamma_i \star x \Longleftrightarrow
    \begin{cases}
    y_i = x_i + 1 \\
    y_j = x_j, \quad j \neq i
    \end{cases}
\end{align}
In this case it is clear that the inverse $\gamma_i^{-1}$ corresponds to subtracting 1 from the $i^{\text{th}}$ index of $x$. Then the symmetric generating set $\Gamma = \{ (\gamma_i, \gamma_i^{-1}) \}_{i \in \{1, 2, \ldots, n\}}$ generates a group which acts on $\mathcal X$.
\end{example}

\begin{figure}
    \centering
    \includegraphics[width=\textwidth]{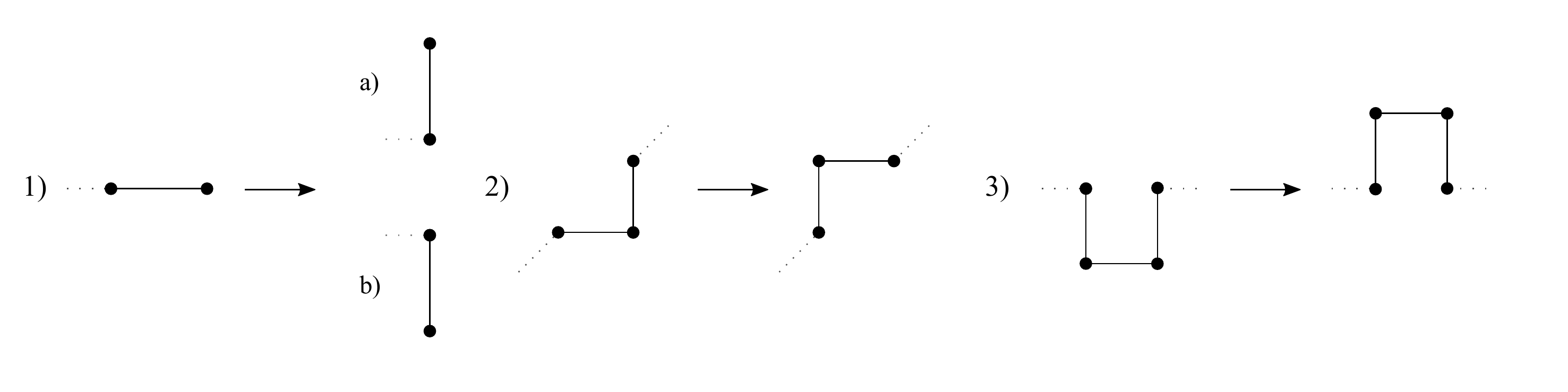}
    \caption{Illustration of the three types of generators for building polymers in protein folding.}
    \label{fig:polymer}
\end{figure}

\begin{example}[Lattice Polymers]
In computational chemistry, one toy model for protein folding is given by lattice polymers, see e.g. \cite{pande1994folding}. Here, one fixes a lattice $L \subset \mathbb{Z}^d$, a positive integer $N$, and considers the space $\mathcal{X}_N$ of length-$N$ walks on $L$, i.e. sequences $( x_0, \cdots, x_N ) \in L^{ N + 1 }$ such that $d ( x_i, x_{i+1} ) = 1$ for all $i$. One also defines a potential function $V: \mathcal{X}_N \to \mathbb{R} \cup \{ \infty \}$, and then seeks to sample from
\begin{align}
    \pi ( x ) \propto \exp \left( - V ( x ) \right) \quad \text{where } x = ( x_0, \cdots, x_N ).
\end{align}
A \textit{polymer} is then defined as a walk which never passes through the same lattice site twice, i.e. $i \neq j \implies x_i \neq x_j$. In this context, it is useful to then think of the potential function $V ( x )$ as being infinite whenever $x$ is a walk which is \textsc{not} a polymer.

In this context, there is a commonly-used group which acts on $\mathcal{X}_N$. This group is simplest to describe by specifying its generators.

The first class of generators are those which arise by moving the end link of a walk, i.e. by taking $x_0$ and moving it to a different neighbour of $x_1$, or taking $x_N$ and moving it to a different neighbour of $x_{N-1}$, see Figure \ref{fig:polymer}, 1).

The second class of generators come from finding subsequences $( x_i, x_{i +1}, x_{i +2} )$ which form an L-shape, and then flipping the orientation of the L as in Figure \ref{fig:polymer}, 2). These are known as `L-flips'.

The third class of generators comes from finding subsequences $( x_i, x_{i +1}, x_{i +2}, x_{i + 3} )$ which make up 3 of the 4 edges of a square, and then rotating these edges around the missing edge, as shown in Figure \ref{fig:polymer}, 3). These are known as `crank-shaft' moves.
\end{example}

We remark that the group generated by these 3 types of move is non-Abelian, and quite complex. However, as in the cyclic group, the order of the elements in the generating set can be determined easily, and is low.

We note for completeness that given a space $\mathcal{X}$, a group $(G, \cdot)$ which acts upon $\mathcal{X}$, and a symmetric generating set $\Gamma$, it is possible to imbue $\mathcal{X}$ with a graphical structure directly via the generators, namely, one declares that $x$ and $y$ are neighbours precisely when there exists a generator $\gamma \in \Gamma$ such that $y = \gamma \star x$. 

We re-emphasize that the benefit of this group-centred formulation over the general discrete space setting is that the space now `looks the same' from all states $x$; no matter where in the space you are, the set of directions in which you are able to move remains the same (in graph-theoretic terms, this is known as \textit{transitivity}). In particular, even though the graph may now have many edges, the `effective' number of edges is now fixed at $| \Gamma |$. This opens us up to constructing more structured stochastic processes with which to sample from distributions defined on $\mathcal{X}$, as notions such as velocity and directionality can now be made sense of in a principled way.

%% file: Inputs/B_Algorithmic_Contributions/GroupsSkewReversible.tex
In this section, we will outline some general principles for designing non-reversible Markov processes of a given invariant measure, on a structured discrete space. A core benefit of non-reversibility is that it allows for the construction of Markov processes which avoid back-tracking behaviour, which is often wasteful.

An observation we have found to be useful is that on discrete spaces with algebraic structure, the benefits of non-reversibility are qualitatively different, depending on whether the generators of the associated group are low-order, or high-order. 

\begin{definition}
Let $(G, \cdot)$ be a group, and let $g \in G \setminus \{ id \}$. We say that $g$ has order $k$ if $g^k = id$, and if $g^j \neq id$ for $0 < j < k$. If there is no such $k$, we say that $g$ has infinite order. We also say that the identity element $id$ has order $1$.
\end{definition}

In particular, if a group has low-order generators (e.g. $\gamma^2 = id$), then there is little benefit in repeatedly applying the same generator, as one swiftly ends up back in a previously-visited state. As such, we might believe that non-backtracking behaviour is best approached by discouraging the repeated use of generators.

In contrast, when the generators have high, or even infinite order, repeatedly applying a generator is taking the process to new states, which allows for persistent motion and exploration. With this in mind, on such spaces, we will try to construct processes which encourage the re-use of generators, when fruitful.

A useful notion in the high-order setting is that of a \textit{reduced generating set}.

\begin{definition}
Let $\Gamma$ be a generating set for a group $G$. We say that $\Gamma_0 \subset \Gamma$ is a reduced generating set if, for all $\gamma \in \Gamma_0$, either $\gamma = \gamma^{-1}$, or $\gamma^{-1} \notin \Gamma_0$.
\end{definition}

Note that a generating set can only be both symmetric and reduced if every element is an involution, i.e. for all $\gamma \in \Gamma_0, \gamma^2 = 1$. The significance of this condition is that it makes it easier to construct processes which are strongly non-reversible, i.e. when it is possible to move in the direction $\gamma$, we can stipulate that moving in direction $-\gamma$ is forbidden. This is a particularly direct way of avoiding back-tracking behaviour.

In the algorithms which follow, we construct non-reversible Markov processes with a particularly tractable form of non-reversibility, known as \textit{skew-reversibility}.

\begin{definition}
Let $\Xc$ be a discrete state space equipped with an involution $S : \Xc \to \Xc$. Let $\pi$ be a probability measure on $\Xc$ such that for all $x \in \Xc$, $\pi ( x ) = \pi ( S ( x ) )$. Let $Q : L^2 ( \pi ) \to L^2 ( \pi )$ be the operator given by $\Qc f ( x ) = f ( S ( x ) )$. 

A Markov Jump Process with generator $\Lc$ is skew-reversible with respect to the measure $\pi$ and the involution $S$ if, for all $f, g \in L^2 ( \pi )$, it holds that 
\begin{align}
    \Eb_\pi \left[ ( \Qc \Lc f ) ( x ) g ( x ) \right] = \Eb_\pi \left[ f ( x ) ( \Qc \Lc g ) ( x ) \right].
\end{align}
By considering $f ( x ) = \mathbf{I} [ x = a ], g ( x ) = \mathbf{I} [ x = b ]$, one can show that this is equivalent to
\begin{align}
    \pi ( x ) \lambda ( x \to y ) &= \pi ( S ( y ) ) \lambda ( S ( y ) \to S ( x ) ) &\quad \text{for all } x, y \in \Xc \\
    \Lambda ( x ) &= \Lambda ( S ( x ) ) &\quad \text{for all } x \in \Xc
\end{align}
\end{definition}

A useful feature of skew-reversibility is that, as with standard reversibility, it provides a checkable, local condition, which ensures that the MJP in question leaves $\pi$ invariant. We will repeatedly use this fact to construct algorithms in the remainder of this section.

In our examples, we will construct processes which are not skew-reversible on the original space $\Xc$, but on an augmented space\footnote{Though note that the discrete Zig-Zag sampler described below suppresses the dependence on $\tau$.} of the form $(x, u, \tau ) \in \Xc \times \Uc \times \{ \pm 1 \}$. Roughly speaking, when the binary variable $\tau$ is equal to $1$, the process will use a certain set of dynamics to move around, and when $\tau = - 1$, the process will run those dynamics in reverse, in a suitable sense. The generality of the construction will mean that we can easily construct skew-reversible Markov processes which admit the correct invariant measure, without requiring additional symmetry assumptions on the state space or target distribution.

%% file: Inputs/B_Algorithmic_Contributions/SAW.tex
The first such process is adapted to the scenario in which the generating set consists of \textit{low-order} elements, that is, for $\gamma \in \Gamma$, we have that $\gamma^k = \text{id}$ for $k \geqslant 2$ a relatively small integer. In many cases (e.g. Bayesian variable selection, binary spin systems, and permutation problems), we can in fact take $k = 2$, and we focus on this case. The heuristic reasoning we apply in this setting is that one should prefer to avoid re-using generators, as if one applies the generator $k$ times, the process has effectively backtracked. As such, we construct an MJP which, over short-to-medium timescales, is able to avoid such behaviour. Due to this property, we term the process the \textit{Tabu sampler}, by analogy with the Tabu search meta-heuristic \cite{glover1998tabu}, which is commonly used in combinatorial optimisation.

The process operates on an extended state space, obtained by augmenting the original space $\mathcal{X}$ with two types of variables. The first is, for each generator $\gamma$, to append an indicator variable $\alpha ( \gamma) \in \{ \pm 1 \}$. The second is a global indicator variable $\tau \in \{ \pm 1 \}$. All of these variables are equipped with the uniform distribution over $\{ \pm 1 \}$.

The behaviour of these variables is as follows: when $\tau = 1$, one is only able to move by using generators such that $\alpha ( \gamma) = 1$, and similarly when $\tau = -1$. Upon making a jump using the generator $\gamma$, one flips the variable $\alpha ( \gamma)$, thus preventing it from being re-used. When the set of available moves becomes too small/unfavourable, the $\tau$ variable flips, and the previously-unavailable generators become available once more. We refer to the path of the process between these $\tau$-flipping events as an `excursion'.

The simplicity of the stationary uniform distribution of the $\alpha(\gamma)$ variables obscures their effective behaviour. In practice, the $\alpha$ variables function as a simple binary memory bank, encoding the successful trajectories of the past: as the algorithm is entering regions of low probability, the typically more desirable backward moves are stored for later use instead, giving the sampler the ability to escape potential wells. The Markov process resulting from running the Tabu sampler is therefore in practice memory-augmented, which is most clearly seen if one resets the $\alpha$ variable mid-run (while still leaving the target invariant). In this case, the sampler loses its sense of direction despite having mixed, and typically spends a long time rebuilding the memory bank before it resumes efficient exploration. Thus the $\alpha$-variable provide a transparent mechanism for avoiding backtracking behaviour, which is necessary for navigating rough energy landscapes. 

\begin{algorithm}\caption{Tabu Sampler for Sampling $\pi ( x ), x \in \mathcal{X} $, when $\gamma^2 = \text{id}$ for all $\gamma \in \Gamma$}\label{alg:saw2}
\begin{enumerate}
\item At $x \in \mathcal{X}, \{ \alpha ( \gamma ) \}_{ \gamma \in \Gamma } \in \{ \pm 1 \}^\Gamma, \tau \in \{ \pm 1 \}$,
    \begin{enumerate}
        \item For $\gamma \in \Gamma$ such that $\alpha ( \gamma ) = \tau$, compute $\lambda ( \gamma; x, \alpha, \tau) = g \left( \frac{ \pi ( \gamma \cdot x ) }{ \pi ( x ) } \right).$
        \item For $\gamma \in \Gamma$ such that $\alpha ( \gamma ) = -\tau$, compute $\lambda ( \gamma; x, \alpha, -\tau) = g \left( \frac{ \pi ( \gamma \cdot x ) }{ \pi ( x ) } \right).$
        \item Compute
            \begin{align}
                \Lambda ( x; \alpha, \tau ) &= \sum_{\gamma \in \Gamma } \lambda ( \gamma; x, \alpha, \tau) \cdot \mathbf{I} [ \alpha ( \gamma ) = \tau ] \\
                \Lambda ( x; \alpha, -\tau ) &= \sum_{\gamma \in \Gamma } \lambda ( \gamma; x, \alpha, - \tau) \cdot \mathbf{I} [ \alpha ( \gamma ) = -\tau ] \\
                \Lambda ( x; \alpha ) &= \max \left( \Lambda (x; \alpha, \tau), \Lambda (x; \alpha, -\tau) \right).
            \end{align}
        \item Sample a waiting time $T \sim \text{Exponential} ( \text{rate } = \Lambda ( x; \alpha ) ) $, and advance time by $T$.
            \begin{enumerate}
                \item With probability $\frac{ \Lambda (x; \alpha, \tau) }{ \Lambda (x; \alpha) }$,
                    \begin{enumerate}
                        \item Sample a new direction $\gamma \in \Gamma$ with probability $\frac{ \lambda ( \gamma; x, \alpha, \tau) \cdot \mathbf{I} [ \alpha ( \gamma ) = \tau ] }{ \Lambda (x; \alpha, \tau) }$.
                        \item Flip the value of $\alpha ( \gamma )$ to $ - \alpha ( \gamma )$.
                        \item Jump to $y = \gamma \cdot x$.
                    \end{enumerate}
                \item With probability $\frac{ \Lambda (x; \alpha) - \Lambda (x; \alpha, \tau) }{ \Lambda (x; \alpha) }$, flip the value of $\tau$ to $-\tau$.
            \end{enumerate}
    \end{enumerate}
\end{enumerate}
\end{algorithm}
From an implementation perspective, the Tabu sampler is essentially the same complexity as the Zanella process; the main cost is still the repeated access to quantities of the form $\frac{ \pi ( \gamma \cdot x ) }{ \pi ( x ) }$. One now has to also maintain the $(\alpha, \tau)$ variables, but this cost is negligible.

\begin{figure}
    \centering
    \includegraphics[width=\textwidth]{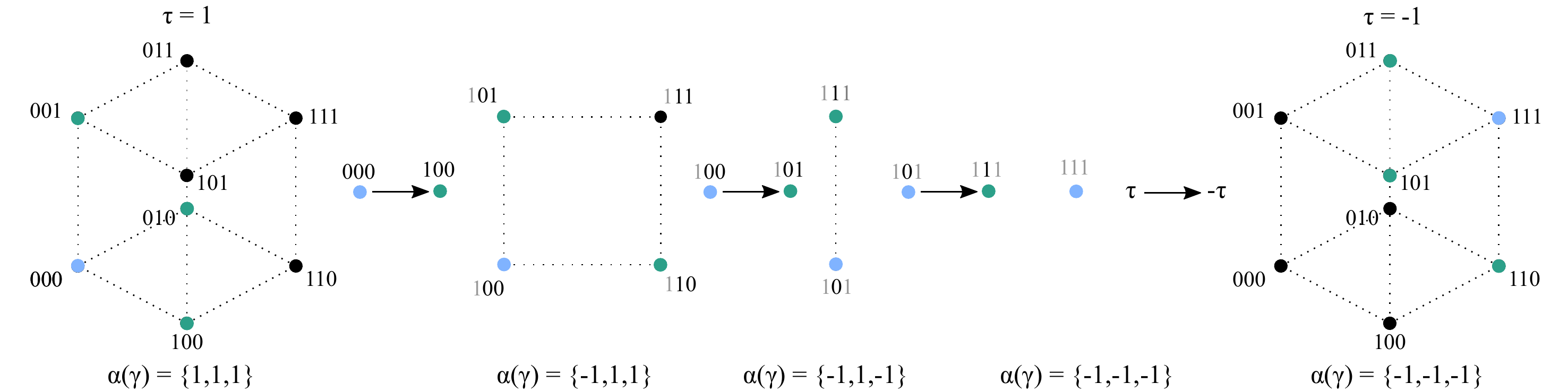}
    \caption{An illustration of the Tabu sampler on the $2\times 2 \times 2$ hypercube. The initial position is illustrated with a light blue dot, the available jumps after applying a generator are coloured light green. For each position, indices available via applications of available generators are coloured black, unavailable locations are light grey. Initially, $\tau = 1$, and all 3 generators are available to use. As the sampler jumps by successively applying generators, the cardinality of the available state-space is halved at each iteration. Furthermore, after 3 jumps the sampler has run out of options, forcing a flip of time (by setting $\tau = -\tau$). This subsequently makes the previously used generators available, and, in this case, all positions become available to the sampler again.}
    \label{fig:dim_reduc}
\end{figure}

An interpretation of the Tabu sampler that can make clear its behaviour is the following. Consider the $|\Gamma|$-dimensional hypercube. For each jump, an entire dimension of the hypercube is removed from the set of accessible points, i.e., after a jump, the side of the cube associated with $\gamma$ is inaccessible until $\tau$ changes direction. In this sense the Tabu sampler performs a dimension reduction at each jump until a reversal of time occurs, after which all previously inaccessible dimensions are made available again. In figure \ref{fig:dim_reduc} we illustrate this behaviour in the simple case of the $2\times 2 \times 2$ hypercube. 

We remark quickly on an interesting `self-tuning' property of the Tabu sampler, namely, the mechanism which allows for $\tau$ to flip. If the total event rate out of the current state, $\Lambda ( x; \alpha, \tau )$ is heavily-dominated by the total event rate out of its mirror state, $\Lambda ( x; \alpha, - \tau )$, then the process is highly likely to flip $\tau$ in order to access the mirror state. In effect, the process is able to discern that flipping $\tau$ would provide a wider range of desirable neighbours to jump to, and is thus somewhat able to adapt to being out of equilibrium in this respect. This also implies that the Tabu sampler will be more effective if the posterior density is multi-modal, as this will encourage diversity in the neighbour set. In numerical experiments, we will estimate the mean excursion as the average number of events of type (d).i which occur before each flip of $\tau$. The realized mean excursion can be interpreted as a proxy for the complexity of the target distribution, as more accepted jumps indicate that the distance between regions of high probability is larger. The situation is one where the Tabu sampler can be expected perform better relative to the Zanella process; see the examples in Section \ref{sec:low_order} for numerical evidence of this. 

The most closely-related existing work we are aware of is the SARDONICS algorithm presented in \cite{hamze2013self}. This is a discrete-time algorithm in which a guided, self-avoiding path of length $k$ is constructed sequentially, the final state of which is then used as a Metropolis-Hastings proposal. While appealing in principle, the algorithm admits some complications; in particular:
\begin{itemize}
    \item Despite constructing a path of length $k$, the proposal ultimately only involves the final state. As such, the self-avoiding path could pass through good potential states which it ultimately has to ignore. As such, the proposal mechanism can be wasteful.
    \item Tuning of $k$, or tuning of the randomisation procedure for $k$ is nontrivial.
    \item As $k$ grows, the complexity of computing the Metropolis-Hastings ratio also grows.
\end{itemize}

In contrast, the Tabu sampler which is presented in this work softens the constraint of being fully self-avoiding, but retains the propensity of being approximately self-avoiding over short-to-medium timescales. As such, from a practical point of view, it appears to present a tractable alternative which retains most of the desirable behaviours of the fully self-avoiding construction.

%% file: Inputs/B_Algorithmic_Contributions/PDMP.tex
Our second class of algorithms is instead adapted to the setting of \textit{high-} or \textit{infinite-order} generators. A reasonable picture to have in mind for this setting is taking $\mathcal{X}$ to be the lattice $\mathbb{Z}^d$, with generators given by the axis-aligned unit vectors. Here, it is less clear that self-avoidance would be beneficial, and it may instead be preferable to \textit{encourage} the re-use of generators, in order to allow for persistent motion across the space.  This heuristic motivates the design of our two \textit{Discrete PDMP} algorithms.

\subsubsection{Discrete Zig-Zag Process}
\input{Inputs/B_Algorithmic_Contributions/ZigZag.tex}

\subsubsection{Discrete Coordinate Sampler}
\input{Inputs/B_Algorithmic_Contributions/Coordinate_Sampler.tex}

\subsection{Related Work}

The most closely-related work we are aware of comes from the literature on non-reversible MCMC. A good starting reference is \cite{diaconis2000analysis}, which presents an in-depth study of a sampler of this form on the state space $\mathcal{X} = \{1, 2, \cdots, N \}$. A number of related constructions are also presented; in particular, the `fiber algorithm' described therein directly motivated the construction of our discrete Zig-Zag Process, which of course bears many similarities to the more recent Zig-Zag Process of \cite{bierkens2019zig}. \cite{chen1999lifting, turitsyn2011irreversible, vucelja2016lifting} are all also closely related in spirit, each focusing on how non-reversibility can be harnessed in order to improve convergence to equilibrium. In the continuous setting, a number of sampling algorithms based around \textit{Piecewise Deterministic Markov Processes} (PDMPs) have also been presented with great success (we recommend \cite{vanetti2017piecewise} for a recent technical overview); these behave in much the same way (hence the name) and served as a large part of the motivation. 

%% file: Inputs/B_Algorithmic_Contributions/ZigZag.tex
While it is possible to construct a version of the Tabu sampler which can handle low-order generators of order greater than two, we found that designing such an algorithm was slightly less natural, and required more tuning choices to be made. We thus opted instead to seek an algorithm which would retain the high-level behaviour of the Tabu sampler, while being more straightforward to tune and implement. This led us to the \textit{discrete Zig-Zag Process} (dZZ).


The dZZ process also operates on an extended state space, though uses a different augmentation. Throughout, we work with a reduced generating set $\Gamma_0$, and for each $\gamma \in \Gamma_0$, we augment the state space with a binary variable $\theta ( \gamma )$, equipped with the uniform distribution over $\{ \pm 1 \}$. The role of $\theta ( \gamma )$ is as follows: at any given time, the process may only use the generator $\gamma$ either forwards (moving from $x$ to $\gamma \cdot x$, when $\theta ( \gamma ) = 1$), or backwards (moving from $x$ to $\gamma^{-1} \cdot x$, when $\theta ( \gamma ) = -1$). Meanwhile, if it were sufficiently beneficial to use the generator $\gamma$ in the opposite direction, then with high probability, the process will flip $\theta ( \gamma )$. In this sense, the process is `self-tuning' in the same fashion as the Tabu sampler.

\begin{algorithm}\caption{Discrete Zig-Zag Process for Sampling $\pi ( x ), x \in \mathcal{X} $}\label{alg:dzz}
\begin{enumerate}
\item At $x \in \mathcal{X}, \theta \in \{ \pm 1 \}^{\Gamma_0}$,
    \begin{enumerate}
        \item For $\gamma \in \Gamma_0$, compute 
            \begin{align}
                \lambda ( x, \gamma; \theta ) &= g \left( \frac{ \pi ( \gamma^{ \theta ( \gamma ) } \cdot x ) }{ \pi ( x ) } \right) \\
                \lambda ( x, \gamma; -\theta ) &= g \left( \frac{ \pi ( \gamma^{ - \theta ( \gamma ) } \cdot x ) }{ \pi ( x ) } \right) \\
                \Lambda ( x, \gamma ) &= \max \left( \lambda ( x, \gamma; \theta ), \lambda ( x, \gamma; -\theta ) \right)
            \end{align}
        \item Compute $ \Lambda ( x ) = \sum_{\gamma \in \Gamma_0} \Lambda ( x, \gamma )$.
        \item Sample a waiting time $T \sim \text{Exponential} ( \text{rate } = \Lambda ( x ) ) $, and advance time by $T$.
        \item Sample a generator $\gamma$ with probability $\frac{ \Lambda ( x, \gamma ) }{ \Lambda ( x) }$, and
        \begin{enumerate}
            \item With probability $\frac{ \lambda ( x, \gamma; \theta ) }{ \Lambda ( x, \gamma ) }$, jump to $y = \gamma^{ \theta ( \gamma ) } \cdot x$.
            \item Otherwise, flip the value of $\theta ( \gamma )$ to $ - \theta ( \gamma )$.
        \end{enumerate}
    \end{enumerate}
\end{enumerate}
\end{algorithm}

A favourable aspect of this process is that it is relatively robust to poorly-scaled targets; if \textit{any} of the available directions are good, then the process is able to sniff them out. The per-iteration cost is comparable to the Zanella process, but the non-reversibility allows for a desirable persistent behaviour.

An added by-product of the dZZ process is that the output of this algorithm can suggest new directions which one could add to the generating set; if one often sees moves where an application of $\gamma_1$ is followed by an application of $\gamma_2$, then one can reasonably augment the generating set to include the move $\gamma_2 \star \gamma_1$. This presents one opportunity for adaptation of these algorithms; we leave exploration of this idea to future work.

%% file: Inputs/B_Algorithmic_Contributions/Coordinate_Sampler.tex
For our third algorithm, we explicitly aim to solve problems in which the group is generated by high-order elements. In this setting, we seek to exhibit persistent behaviour across the space, i.e. if moving from $x$ to $\gamma \cdot x$ is successful, then we will attempt to make additional moves to $\gamma^2 \cdot x, \gamma^3 \cdot x$, and so on.

As in the previous two cases, we operate on an extended state space, though the interpretation is now somewhat different. We first add in a variable $v$, taking values in our symmetric generating set $\Gamma$, which is drawn according to some symmetric distribution $\psi$ (i.e. such that $\psi ( v ) = \psi ( v^{-1} )$. We then include a `direction of time' variable $\tau$, which is equipped with the uniform distribution on $\{ \pm 1 \}$. In effect, $v$ behaves as a velocity, and $\tau$ dictates whether to follow the velocity forwards or backwards in time.

\begin{algorithm}\caption{Discrete Coordinate Sampler for Sampling $\pi ( x ), x \in \mathcal{X} $, $v \sim \psi ( v )$}\label{alg:dcs}
\begin{enumerate}
\item At $x \in \mathcal{X}, v \in \Gamma, \tau \in \{ \pm 1 \}$,
    \begin{enumerate}
        \item Compute
            \begin{align}
                \delta (x, v, \tau) &= g \left( \frac{ \pi ( v^\tau \cdot x ) }{ \pi ( x ) } \right) \\
                \delta (x, v, - \tau) &= g \left( \frac{ \pi ( v^{-\tau} \cdot x ) }{ \pi ( x ) } \right) \\
                \Delta ( x, v ) &= \max \left( \delta (x, v, \tau), \delta (x, v, -\tau) \right).
            \end{align}
        \item Sample a waiting time $T \sim \text{Exponential} ( \text{rate } = \Delta ( x, v ) ) $, and advance time by $T$.
            \begin{enumerate}
                \item With probability $\frac{ \delta (x, v, \tau) }{ \Delta (x, v ) }$, jump to $y = v^\tau \cdot x$.
                \item Otherwise, with probability $\frac{ \Delta (x, v ) - \delta (x, v, \tau) }{ \Delta (x, v, \tau) }$, sample a new velocity $w$ according to
                    \begin{align}
                        Q ( w | x, \tau ) &= \frac{ \psi ( w ) \rho ( x, w, \tau ) }{ Z ( x ) } \\
                        Z ( x ) &= \frac{1}{2} \sum_{ \tau \in  \{ \pm 1 \} } \sum_{\gamma \in \Gamma } \psi ( \gamma )  \rho ( x, \gamma, \tau ),
                    \end{align}
                where
                    \begin{align}
                        \rho ( x, v, \tau ) = \left[ \delta ( x, v, -\tau ) - \delta ( x, v, \tau ) \right]_+
                    \end{align}
                set the value of $v$ to $w$, and flip the value of $\tau$ to $-\tau$.
            \end{enumerate}
    \end{enumerate}
\end{enumerate}
\end{algorithm}

Broadly, the walk attempts to follow the velocity in the direction of time $\tau$. When the walk is heading towards regions of higher probability, $\delta (x, v, -\tau)$ will be equal to $0$, and thus the walk will continue make moves in the direction $v^\tau$. Once the walk has started to head towards regions of lower probability, the $\delta (x, v, -\tau)$ term will begin to dominate, and the walk will instead try to modify its own velocity, and pursue a new direction.

We note that relative to the discrete Zig-Zag Process, the computational complexity of making a single forwards jump with the discrete Coordinate Sampler is much cheaper; in particular, it is independent of the size of the generating set. This represents a substantial benefit in certain high-dimensional scenarios. The cost is then that some robustness is lost; as the discrete Zig-Zag process is always able to look in multiple directions, it should be able to adapt to changing `curvature' more gracefully. In contrast, for `anisotropic' target distributions, the discrete Coordinate Sampler will likely have to resample its velocity variable quite frequently, which could offset the reduced per-iteration cost.

We caution quickly that there is an odd pathology which can cause the sampler to become reducible when sampling from highly-symmetric target distributions. In particular, if there is a subset $S \subset \mathcal{X}$ and a generator $v$ such that $\delta ( x, v, \tau ) = \delta ( x, v, - \tau )$ for all $x \in S$, then whenever a velocity jump event is experienced in $S$, the newly-resampled velocity cannot be equal to $v$, and this can lead to reducibility under certain circumstances. This is often not a problem - in particular, if the other velocities can allow the process to exit the set $S$, then the problem typically vanishes - but for certain initialisations, this can cause undesirable behaviour. As such, we generally advise that these algorithms should be initialised \textit{away} from points with some symmetry to them.

%% file: Inputs/D_Numerical_Examples/Implementation.tex
As discussed previously, a MJP on a discrete space can be implemented exactly without discretization error, as no numerical integration is necessary. To estimate statistical quantities $\mathbf E_\pi [f(x)]$ for $f: \mathcal X \rightarrow \mathbb R$ from a realisation of a process $(X)_{t \in [0,T]}$, two approaches are available. By ergodicity, the empirical time-average approaches the true expectation in the limit:
\begin{align}
    \lim_{T\rightarrow \infty} \frac{1}{T} \int_{[0,T]} f(X_s) ds = \mathbf E_\pi \left[ f(x) \right],
\end{align}
and the integral on the left-hand side may be calculated explicitly as
\begin{align}
    \frac{1}{T} \int_{[0,T]} f(X_s) ds = \sum_{k \geq 1} \frac{\tau_k - \tau_{k-1}}{T} f(X_{\tau_k}),
\end{align}
where the sum ranges over each event-time and $X_{{\tau_k}}$ is the value of the process just after the $k$'th event. 
Alternatively, a thinning procedure can be applied to the process. In this case, a thinning interval $0 < \vartheta < T$ is chosen such that $\frac{T}{\vartheta}$ is an integer, and at each of these thinning times the process state is stored. Expectations are then calculated simply via the empirical average
\begin{align}
    \mathbf E_\pi \left[ f(x) \right] \approxeq \frac{\vartheta}{T}\sum_{i=0}^{\frac{T}{\vartheta}} f(X_{i \cdot \vartheta}).
\end{align}
The resulting difference in the estimates from using the thinned samples is in practice completely negligible, however, the thinning procedure allows one to explicitly define the process run-time $T$ when pre-allocating storage for each thinned sample, while the discrete time integral-approach requires pre-allocating storage for each event-time, which implies that the final time $T$ is random, or that unused excess storage is necessary. For these reasons, we apply the thinning method to generate samples while the algorithm is running. After an initial trial run, we set the thinning rate $\vartheta$ to approximately be equal the mean event-time after the process has reached stationarity. In other words, in this case we on average expect a single event to have occurred for each thinned sample. 

%% file: Inputs/D_Numerical_Examples/LocallyBalancing.tex
In \cite{zanella2019informed} it is shown that balancing functions are Peskun-optimal \cite{peskun1973optimum} weighting functions. Although the author establishes that, in the particular case of independent Bernoulli variables, the Barker balancing function is optimal, there is still far from a complete theory concerning the optimal choice of $g$ for general targets $\pi(x)$. 

To explore these issues, we consider the \textit{embedded jump chain} $(\hat{x}_k)_{k \geq 1}$ of the Zanella process, which is the discrete-time Markov chain consisting of values of the process evaluated after each event-time $\tau_k$, i.e. $\hat{x}_k = X_{\tau_k}$, where $\tau_0 = 0$. The invariant distribution of the jump chain (hereafter, the `jump measure') is determined by the choice of balancing function as
\begin{align}
    \pi_g^J(x) \propto \pi(x) \Lambda (x) = \pi(x) \left( \sum_{y \in \partial x} \lambda ( x \to y ) \right) = \pi(x) \left( \sum_{y \in \partial x} g \left( \frac{ \pi(y) }{ \pi (x) } \right) \right).
\end{align}

By studying the effects of the balancing function on the jump measure, we can seek to understand its effects on the mixing of the underlying Markov process. 

One approach which may prove insightful here is to consider how the choice of $g$ affects the metastability of the jump chain. In particular, if the jump measure exhibits lower energy barriers between modes, we should expect more desirable mixing behaviour. 

\begin{figure}
    \centering
    \includegraphics[width=\textwidth]{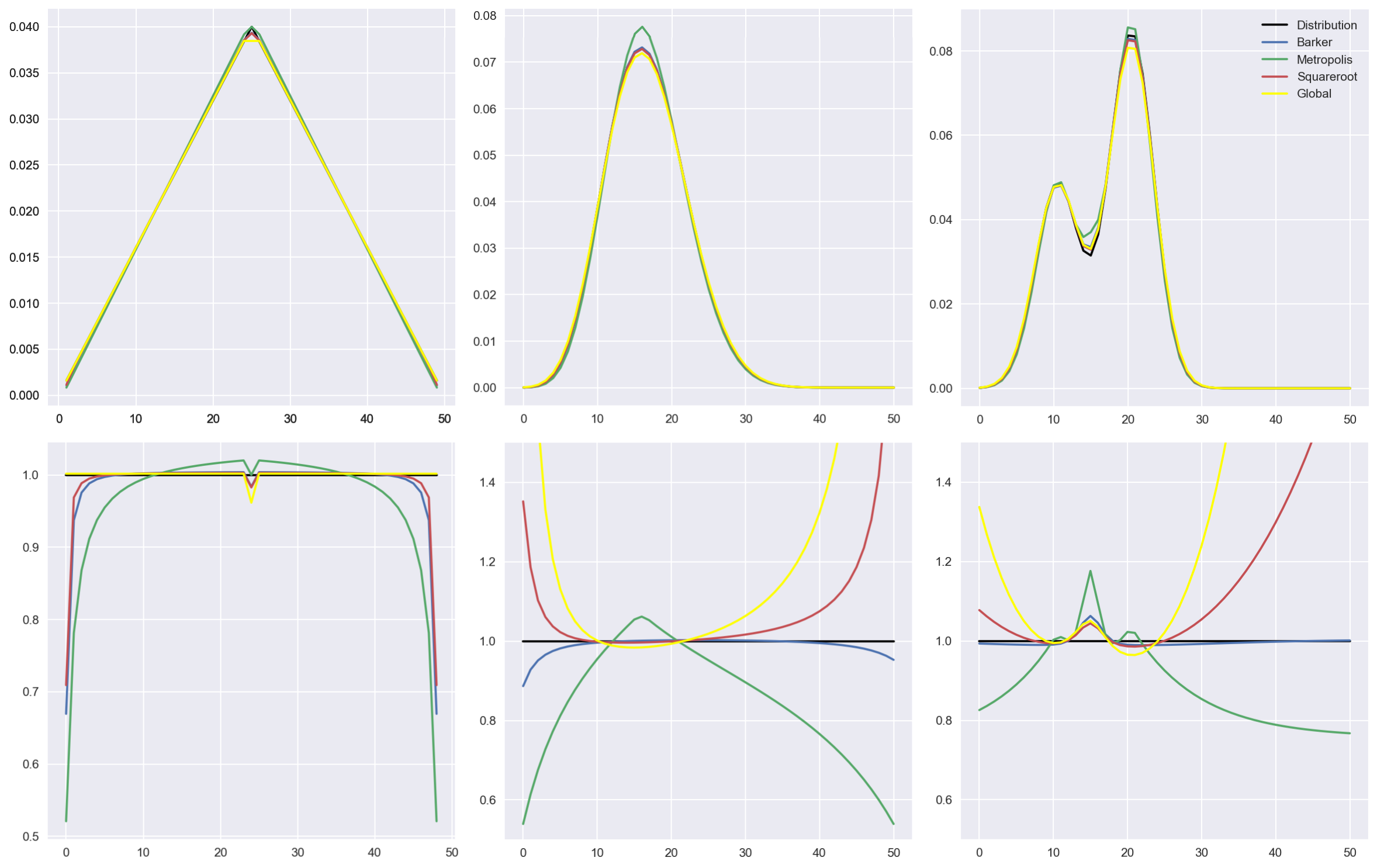}
    \caption{Upper row: Three distributions (Triangle, Beta-Binomial(10, 20) and a marginal of a mixture of lattice Gaussians) and their jump-chain invariant distributions for the three admissible balancing functions and the global balancing function. Lower row: Relative ratio of the jump-distribution to the invariant, $\pi_g^J(x)/\pi(x)$.}
    \label{fig:balancers}
\end{figure}

A more concrete way to probe this relationship is to study the ratio of the target distribution $\pi$ to the jump measure $\pi^J$, as it provides a sense of where the jump process places emphasis relative to the target. We present here some simple one-dimensional examples on $\mathcal X = [50]$, where the neighbourhood structure is defined by setting $\partial x = \{ x - 1, x + 1 \}$. In Figure \ref{fig:balancers}, we plot three distributions of increasing complexity, and the corresponding jump measures for the different balancing functions. We also test the \textit{non-balanced} (or `globally-balanced') weighting function $g(t) = t$ to contrast its behaviour. 

One heuristic for desirable behaviour is for the jump measure to be remain close to $\pi$, as larger deviations require more effort from the continuous time process to correct for. This heuristic argument is related to what is put forward in \cite[Section 2.1]{zanella2019informed} in favour of balancing functions. 

It is clear from Figure \ref{fig:balancers} that an increased complexity of the target distribution emphasises the need for balanced weighting functions, but even within the class of balancing functions, one can observe significant differences in behaviour. For the Metropolis balancing function, modes are over-emphasized in the jump measure, while low-probability regions are visited significantly less often. The square root is generally more robust around modes, but in contrast to the Metropolis balancing function, puts significantly higher probability on extreme regions. The Barker function seems in general to balance the best features of Metropolis and square root, with reasonably stable behaviour around peaks and troughs, and significantly less erratic tail behaviour. The non-balanced function performs better than its local counterparts for relatively flat targets, but for more interesting targets performs exactly opposite to what is desired. 

%


%% file: Inputs/D_Numerical_Examples/LowOrder.tex
We here present five examples that cover a wide swathe of discrete models applied in statistics, machine learning and physics. For each of the examples it is the case that the generators are of order 2. By the arguments presented in \cite{zanella2019informed}, we only compare our samplers to the Zanella process, as the superior performance and theoretical underpinnings of locally-balanced samplers implies that it is less relevant to compare our algorithms to either globally-balanced samplers, such as the Hamming Ball sampler \cite{titsias2017hamming}, or random-walk schemes. In Table \ref{tab:ess_comparison}, we provide a short summary of the performance of the Tabu sampler relative to the Zanella process in terms of the ratio of effective sample size per second (ESS/s)\footnote{Care has been taken to ensure that the implementation of the samplers is fully comparable in terms of computational resources spent per event.}, with higher being better for the Tabu sampler. We also note the mean excursion, which we defined above as average length of the self-avoiding walk. As mentioned, the mean excursion provides a good indication of the complexity of the target distribution, and we in general expect the performance of the Tabu sampler relative to processes that allow backtracking to increase as the mean excursion goes up. For each experiment, we ran the samplers 5 times and averaged over each run. Furthermore, for the experiments involving simulated data, we also re-initialized the data to increase the accuracy of the comparison.

\begin{table}[h]
\centering
\begin{tabular}{@{}llll@{}}
\toprule
                             & \textbf{Target Statistic}    & \textbf{Mean Excursion} & \multicolumn{1}{l}{\textbf{Tabu/Zanella}} \\ \midrule
Bayesian Variable Selection  & Number of parameters & 6.0           &   1.57                                       \\
Conditional Permutation Test & Hamming distance from mode   & 12.8           & 1.16                                         \\
Sherrington-Kirkpatrick Model      & Energy               & 83.4         & 79.89                                         \\
Log-Submodular Point Process & Sensor count         & 31.8         & 24.48                                        \\
Determinantal Point Process  & Number of points               & 11.0           &    4.98                                      \\ \bottomrule
\end{tabular}
\caption{Summary of performance for target distributions with order-2 generators over 5 runs. 'Mean Excursion' is the average number of events between a time-reversal of $\tau$. 'Tabu sampler/Zanella' is the ratio of ESS/s for the Tabu sampler relative to the Zanella process after burn-in, discarding the first 20\% of 100.000 samples. Each ESS calculation is carried out with autocorrelation lags up to order 3000.}
\label{tab:ess_comparison}
\end{table}

%% file: Inputs/D_Numerical_Examples/BayVar.tex
The probabilistic selection of covariates in problems such as regression has long been an active part of theoretical and applied Bayesian statistical research \cite{mitchell1988bayesian, george1997approaches, lee2003gene}. While efficient algorithms exist for fitting many of the most popular Bayesian models, the hierarchical framework allows for complex interactions that can make posterior exploration very difficult. 
We analyze here a hierarchical Bayesian model presented in \cite{schafer2013sequential}, which postulates that the $m$-dimensional observation vector $y$ in a linear regression framework is observed through
\begin{align}
    y \hspace{4pt} | \hspace{4pt} \beta, x, \sigma^2, Z \sim \mathcal{N}(Z I^x \beta, \sigma^2 I_m),
\end{align}
with covariates $Z \in \mathbb R^{m \times n}$, parameters $\beta \in \mathbb R^n$ and the matrix of active covariates $I^x = I_n x$, with the binary inclusion vector $x \in \mathcal{X} = \{0,1\}^n$ our target variable. The group action here is equivalent to the one presented for spin glasses in Example \ref{spinglass_action}, with the generator in this case instead given by picking a single entry in $x$, $x_i$, and setting $x_i' = 1 - x_i$. To achieve a closed form marginal posterior that is independent of $(\beta, \sigma^2)$
\begin{align}
    \pi(x \vert Z, y) = \int \pi(x \vert Z, y, \beta, \sigma^2) d(\beta, \sigma^2),
\end{align}
we pick conjugate priors 
\begin{align*}
    p(\beta \vert \sigma^2, x) &= \mathcal{N}(0, v^2 \sigma^2 I^x), \quad p(\sigma^2) = \text{I}\Gamma \left( \frac{w}{2}, \frac{\lambda w}{2} \right), \quad
    p(x) = U(\mathcal{X}),
\end{align*}
with I$\Gamma$ the inverse-gamma distribution and $U(\mathcal{X})$ the uniform law on $\mathcal X$. Setting the hyper-parameter vector $(w, v, \lambda)$ as in \cite{george1997approaches}, we use the Concrete Compressive Strength dataset, originally analysed in \cite{yeh1998modeling}, which includes 8 covariates used to explain the compressive strength of concrete, as measured in giga-Pascals. The dataset is augmented with a constant column of ones, five logarithmic variables, and first-order interactions of the log-transformed and initial covariates for a total of $n = 92$ parameters and $m = 1030$ observations in the saturated model. We do not enforce main effects restrictions, which implies that interaction terms can be included independently of whether baseline covariates are included in the model or not. To have manageable rate sizes, the Barker balancing function $g(t) = \frac{t}{1+t}$ is actually needed in this case, as the change in probability from changing the variable subset can be substantial. 
The goal of sampling $\pi(x)$ is to derive the marginal inclusion probability $\rho_i = \mathbf{E}_\pi[\mathbf{1}_{x_i = 1}(x)]$ of each variable in the model, rather than just MAP estimates as is commonly done. The resulting inclusion probabilities from the Tabu sampler or the Zanella process are virtually indistinguishable from the ones estimated in \cite[Figure 5]{schafer2013sequential}, in comparison with the huge variability observed for the discrete-time adaptive MCMC and standard MCMC samplers there. Nonetheless, the total number of evaluations of the posterior density is higher for the Zanella process and the Tabu sampler compared to the tailor-made SMC procedure applied by \cite{schafer2013sequential}, which only evaluates the density once per iteration.

To evaluate the effective sample size, we decide against using the energy, as the posterior is dominated by a single configuration, and this mode is often revisited. Rather, we use the number of active variables as the test statistic. It is noteworthy here that there seems to be limited benefit available from using the Tabu sampler for this dataset, as the posterior appears to be sufficiently well-behaved that the self-avoiding behaviour is largely unnecessary. As such, using locally-balanced proposals allows for the posterior to be explored properly and efficiently. In fact, across some time-scales, the self-avoiding property of the Tabu sampler can actually impede the mixing of the chain, as the sampler can corner itself and be forced to move in a bad direction, recall Figure \ref{fig:dim_reduc} where the sampler ran out of options after 3 jumps. This stands in contrast to the Zanella process, which is always able to revert to a previous configuration. Nonetheless, the slightly better performance of the Tabu sampler over repeated runs can reassure us that self-avoidance does not lead to worse performance overall. 

%% file: Inputs/D_Numerical_Examples/CPerm.tex
We here consider a version of the matching problem already described in Example \ref{example:linkage}. Let $S = \{1,2,\ldots, n\}$ and consider the space $\mathcal{X} = \Sigma_n$ of permutations of $S$. The goal is to explore likely orderings of $S$ that have high likelihood under our model, which we now describe. For all integer combinations $(i, j) \in S\times S$ we associate a likelihood of that particular pairing given by $\omega_{i, j}$, where $\omega_{i,j} \sim \log \mathcal{N}(0, \sigma^2)$ independently. For any permutation $x \in \mathcal X$ we define the target density as
\begin{align}
    \pi(x) = \frac{1}{Z}\prod_{i=1}^n \omega_{i, x_i},
\end{align}
with $Z = \sum_{x \in \mathcal X}\prod_{i=1}^n \omega_{i, x_i}$ the normalizing constant. For large $\sigma^2$ the log-normal has quite heavy tails, and thus the distribution will be dominated by some pairings being much more likely than others, giving rise to multimodality. However, in comparison with for example the much more narrow distribution of the correlation coefficients $J_{ij}$ in the Sherrington-Kirkpatrick model below, only a few steps are on average needed to travel between close modes. In practice, for the most difficult i.i.d. case considered in \cite{zanella2019informed} where $n = 500$ and $\sigma^2 = 5$, the average excursion length is around 13, which is comparable to what was observed in the determinantal point process example of Section \ref{seq:dpp}. Overall, the Tabu sampler performs similarly to the Zanella process, indicating that there is little benefit of avoiding backtracking behaviour when the distribution is very peaked at the modes. 

%% file: Inputs/D_Numerical_Examples/SpinGlass.tex
Spin glasses are a class of models of magnets with competing ferromagnetic and antiferromagnetic interactions, and which are widely studied in statistical physics, in particular condensed matter, but also applied in diverse fields such as protein folding \cite{bryngelson1987spin}, neuroscience \cite{fuhs2006spin} and spatial economics \cite{krugman1994complex}. As in example \ref{spinglass_action}, consider a 2-dimensional lattice $V = L^2_n$ with side-length $n$, where each vertex is inhabited by a binary spin $x_i \in \{ \pm 1\}$. We let $|V| \equiv n^2$ be the overall cardinality of the graph.

We will analyze the Sherrington-Kirkpatrick (SK) model, a lattice-wide model with random interactions on V, implying that the graph $(V,E)$ is complete, i.e. fully connected. The log-probability, or negative Hamiltonian $H$, of the target is
\begin{align}
    \log\pi(x) \equiv -\beta H(x) = \frac 1 n \sum_{i=1}^n\sum_{j=1}^n J_{ij} x_i x_j +  h \sum_{l=1}^n x_l = \frac 1 n \sum_{i=1}^n x_i \sum_{j \neq i} J_{ij} x_j + h\sum_{l=1}^n x_l + \frac{J_{ll}}{nh},
\end{align}
where $J_{ij} \sim \mathcal{N}(0, \beta^2 (2n)^{-1})$ if $ i \neq j$ and $0$ otherwise, and $h > 0$. In both cases, we simply absorb the inverse temperature $\beta$ into the likelihood constant $h$ and the correlation coefficient.  It follows that there are $|V|^2$ interaction terms, so the model scales in storage costs at order $O(n^4)$. The SK model generalizes many of the common spin-glass models: If $J_{ij} = c$, a constant, for all $(i,j) \in L^2_n$, the model reduces to the Curie-Weiss model. On the other hand, if the model is local in the sense that two vertices $(i,j)$ and $(l,m)$ only interact when $|i-l| + |j-m| = 1$, the model reduces to the Edwards-Anderson model. Finally, if both simplifications are assumed, we revert to the classical Ising model. 

With the generator flipping the $i^{\text{th}}$ vertex denoted by $\gamma_i$, we can exploit the sum-structure of the probability distribution when calculating the rates. The jump rates can be calculated via
\begin{align}
    \lambda (\gamma_i; x) = g \left( \frac{\pi(\gamma_i x)}{\pi( x)} \right) = g \left( \exp \left \{ H(\gamma_i x) - H( x) \right\} \right) = g \left( \exp \left\{ \frac{4}{n} x_i \left( \sum_{j=1}^n J_{ij}x_j + \frac{1}{2} h \right) \right\} \right).
\end{align}
More importantly, the update of vertex $l$ given a previous update at $i$ is simply given by $\partial_l H(\gamma_i x) = -\frac{8}{n} J_{il} x_i x_l$, which significantly speeds up computations. We evaluate the model with $n = 100$, $\beta = 10$ and $h = 0.1$. For initialisation, we pick $x_0 = \{1,1,1, \ldots, 1, 1\}$. The highly multi-modal nature of the Hamiltonian is a situation where the Tabu sampler can be expected to do well, as proper exploration is contingent upon leaving potential wells, which should be easier as generators leading back towards the well are removed from $\alpha(\gamma)$. 

\begin{figure}
    \centering
    \includegraphics[width=\textwidth]{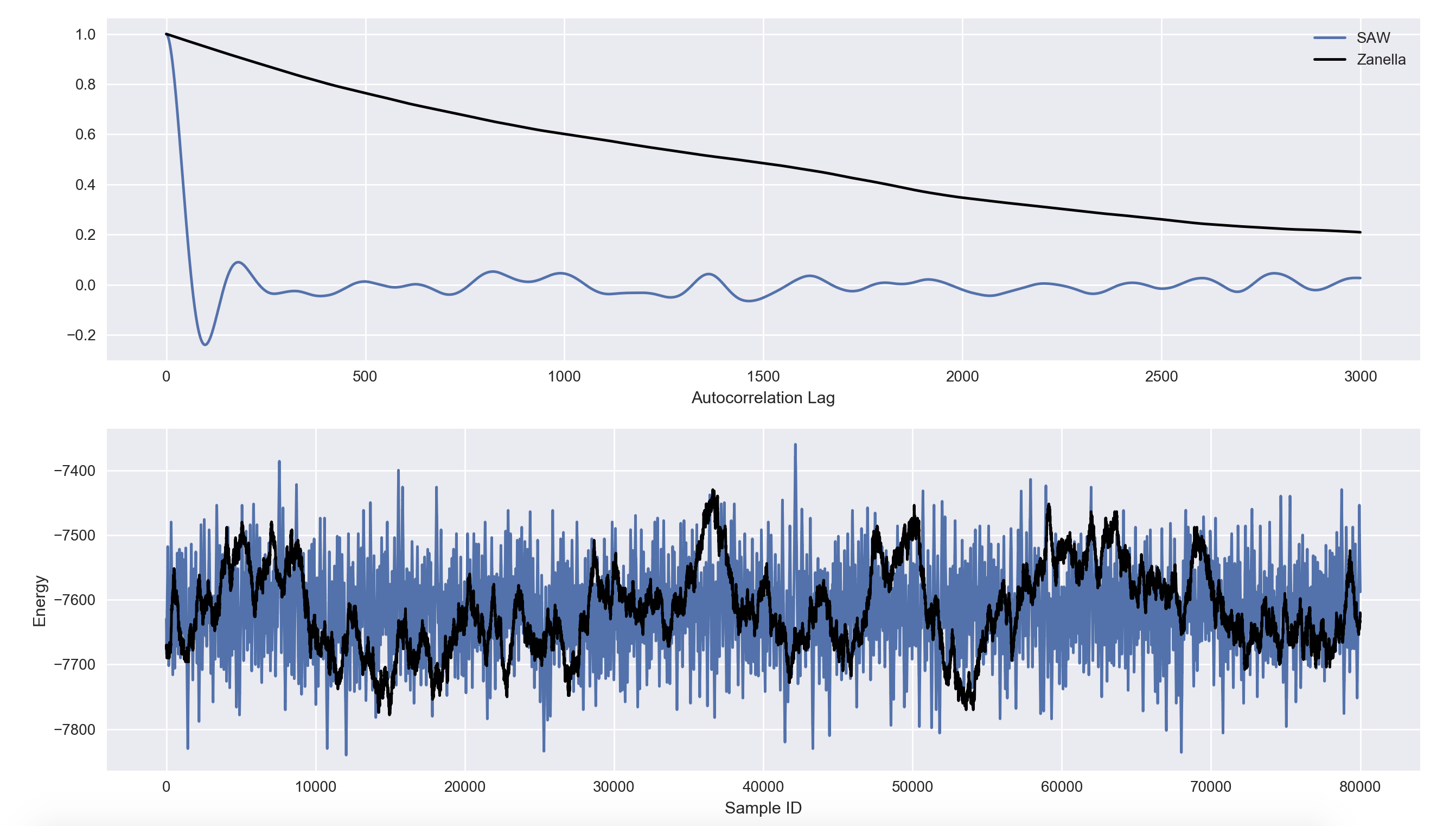}
    \caption{Autocorrelation function and energy traceplot for 80.000 burned-in samples drawn from the $100 \times 100$ Sherrington-Kirkpatrick Hamiltonian via the Tabu sampler and Zanella process. }
    \label{fig:s-k_traceplot}
\end{figure}

In practice, we observe that for our spin-system with 10,000 spins, the average excursion length of the Tabu sampler is 83, or nearly 1\% of the possible spins. In comparison with the four other examples, this is the highest observed excursion length, indicating that modes are distantly spaced for the Sherrington-Kirkpatrick model, in comparison for example with the conditional permutation test of Section \ref{seq:cpt}. We conjecture in general that the Tabu sampler will perform better than other neighbourhood-based samplers the further spaced modes are in terms of group actions required to reach a neighbourhood of a mode; we postpone theoretical analysis to future work. In terms of ESS/s, the Tabu sampler performs close to two order of magnitudes better than the Zanella process, which already in the discrete-time experiments of \cite{zanella2019informed} was shown to perform 40-50 times better for the same metric, in the much simpler case of a ferromagnetic 2D Ising model, than random walk proposals, the HB sampler \cite{titsias2017hamming} and the D-HMC sampler \cite{pakman2013auxiliary}. In Figure \ref{fig:s-k_traceplot}, we display the autocorrelation function and the trace plot of the energy for the two samplers. 

%% file: Inputs/D_Numerical_Examples/SubPoint.tex
Modular functions have found broad application within machine learning and combinatorial optimization (see for example \cite{bach2019submodular}) in recent years, and and provide a cohesive framework for modelling repulsion and regularity of subsets of points in space. For some integer $m$, consider the set $\mathcal{X} = 2^m$. We say a function $f: \mathcal{X} \rightarrow \mathbb{R}$ is submodular if for any sets $S, T$ such that $S \subseteq T$ and any $i \in \mathcal{X} \setminus T$,
\begin{align}
    f(S \cup i) - f(S) \geq f(T \cup i) - f(T),
\end{align}
that is, there are diminishing marginal returns from making sets larger. To define a probability distribution, we simply let up to proportionality the probability of a given set $S$ be $\pi(S) \propto \exp \{ f(S) \}.$ In the following two examples we consider variations on these kinds of distributions. 

We first consider a variation on the facility location problem \cite{charikar1999improved, jain2001approximation} for internet access points. Let $\mathbb{K}$ denote a non-convex polygon in $\mathbb R^2$. Inside $\mathbb K$ we have a set of $m$ equidistant points consisting of $(v_l)_{1 \leq l \leq m}$, from hereon known as access points. Also in $\mathbb K$, there are $n$ individual users $(u_j)_{1 \leq j \leq n}$ placed randomly according to a Gaussian distribution centered at the centroid of $\mathbb K$. For each user $u_j$ we calculate the utility available from using the access point $v_i$ as $\Upsilon_{ij} = \exp \{ -\kappa \Vert u_j - v_i\Vert^2 \}$, where $\kappa$ is a parameter representing the physical signal decay, the distance is contained in the convex hull of $\mathbb K$. Consider the value function $h$
\begin{align}
    h(S) = \sum_{j=1}^n \max_{i \in S} \Upsilon_{ij};
\end{align}
the function returns the maximum value that can be extracted for all users given they optimally use the closest access point available to them. For installation costs, we have $g_1(S) = -\lambda |S|$ for some fixed cost-parameter $\lambda > 0$. The probability distribution given by
\begin{align}
    \pi(S) \propto \exp \big \{ h(S) - g_1 (S) \big \}
\end{align}
is then a log-submodular probability distribution. To incorporate capacity constraints, we introduce $C(S, j) = \argmax_{i \in S} \Upsilon_{ij}$, the choice of sensor by user $j$, $B(S, i) = \sum_{j=1}^n \mathbf{I}[C(S, j) = i]$, the total number of users of sensor $i$, and the capacity cost function
\begin{align}
    g_1(S) = \psi \sum_{i=1}^m \max \{0, B(S, i) - \Psi \}
\end{align}
with $\psi > 0$ modelling user response to exceeded capacity and $\Psi \in \mathbb N$ the maximum capacity before bandwidth declines below some acceptable threshold. However, although the function $f' = h - g_1 - g_2$ is not submodular, it provides an example of a capacity-constrained facility location problem, and thus an interesting benchmark case, given that it incorporates significant shifts in probability when capacities are reached for access points. \\

\begin{figure}
    \centering
    \includegraphics[width=\textwidth]{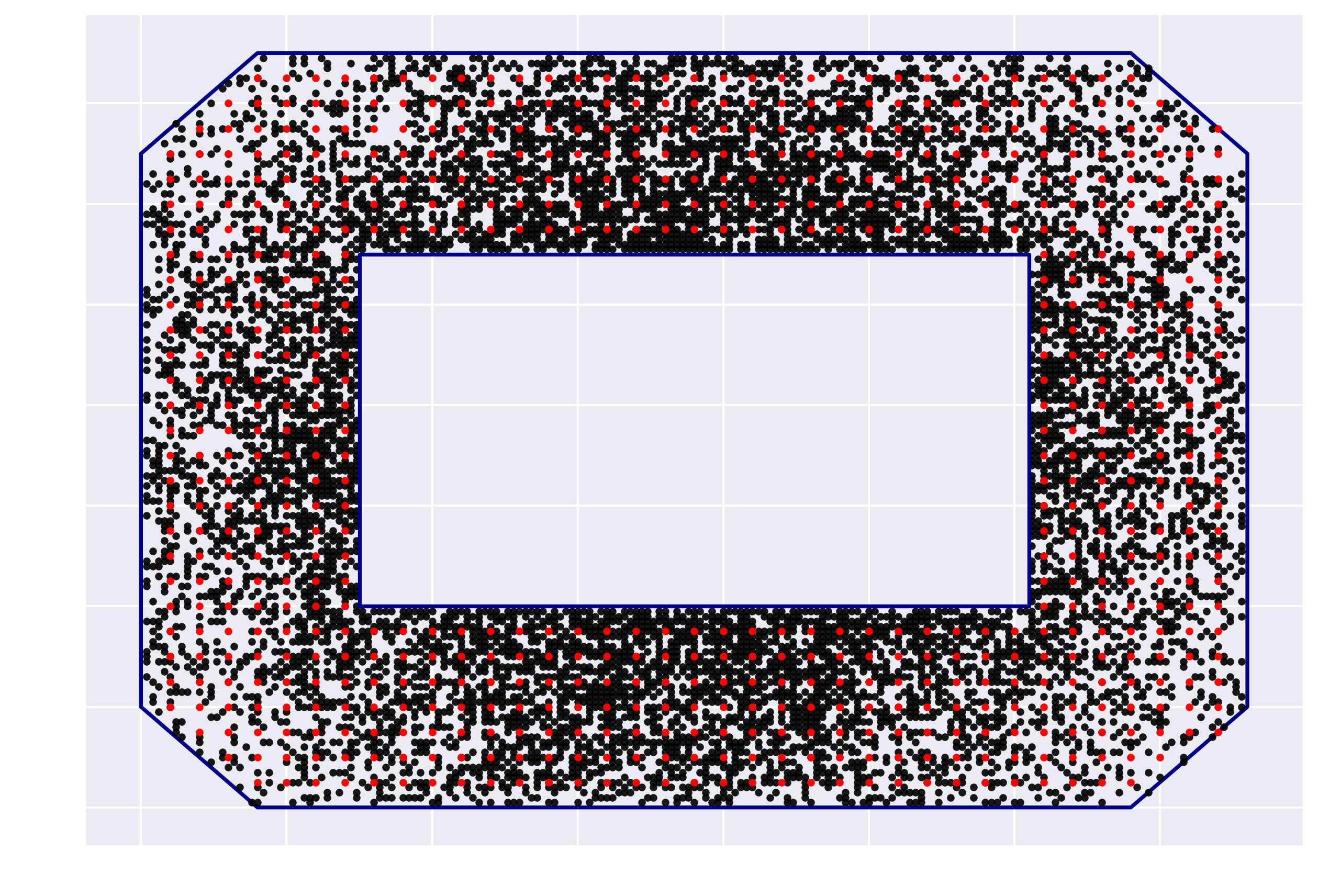}
    \caption{Toy-model of a stadium. Each black dot corresponds to a spectator, and each red dot corresponds to a sensor. }
    \label{fig:stadium}
\end{figure}

As an illustrative example, we consider the case of providing access to randomly placed users where $\mathbb K$ is given as in Figure \ref{fig:stadium}, a model football stadium. The access point grid considers a vertical and horizontal spacing of 5 for a total of $m = 704$ access points, and a random number of spectators are drawn inside the stadium boundaries with higher propensity of being along the lateral sides of the field, the expected number of spectators is 8750. Assuming that each access point provides $100$ $mbit/s$ download and upload rates, we set the capacity constraint to $\Psi = 25$ users, and $\lambda = \psi = 1$, which corresponds to the complete loss of utility for a single optimally placed user for spectator connected user in excess of access point capacity. For the utility gained, we use an adjustment factor $\kappa = n$. To assess the capacity for the samplers to provide suitably different configurations of sensors, we use the number of sensors as the target statistic. The observed mean excursion in this example is just below 32 for the Tabu sampler, indicating significant complexity in the distribution, and the improvement in effective sample size relative to the Zanella process is more than an order of magnitude. We note that the capacity constraint is rarely breached for this particular set of parameters, and the resulting log sub-modular density without capacity constraints exhibit similar results as the constrained one, with just a slight decline in the relative performance of the Tabu sampler.

%% file: Inputs/D_Numerical_Examples/DetPoint.tex
Similar to the facility location example above, we here consider sampling subsets $S$ of elements of a pre-fixed set of items $[m] = \{1,2,\ldots, m\}$ for some positive integer $m$. A determinantal point process (DPP), generically denoted $\mathcal P$, is a point process on $[m]$ taking values in $\mathcal X = 2^m$, we call each such subset a point configuration, see \cite{kulesza2012determinantal} for a very comprehensive overview. The defining feature of DPPs is that there is some real, positive semi-definite matrix $K$, denoted the marginal kernel, that captures a sense of 'closeness' or similarity of the different items. We then say a point process $\mathcal P$ is a DPP if whenever $X\in \mathcal X$ is some random subset drawn from $\mathcal P$ and $S \subseteq \mathcal \{1,2,\ldots, m\}$,
\begin{align}
    \mathbf P (S \subseteq X) = \det K_{S},
\end{align}
where $\det \cdot$ denotes the determinant, and $K_S$ is the restriction matrix of $K$ with corresponding rows and columns corresponding to the active indices. In this case, the probability of including any single entry $1 \leq j \leq m$ in $X$ is just equal to the $j^{\text{th}}$ diagonal entry of $K$, so in particular for any two indices $i, j$, we have
\begin{align}
    \mathbf P (\{i,j\} \subseteq X) = \det K_{\{i,j\}} = K_{i,i}K_{j,j} - K_{i,j}^2,
\end{align}
such that the probability is decreasing the closer the two points are in terms of the marginal kernel. By this property, DPPs explicitly model repulsion of points. A DPP is also an example of a log-submodular probability distribution, as the determinant is always increasing in the number of points. For our purposes, it is easier to work with an explicit form of DPPs known as L-ensembles. Let $L$ be the real $m\times m$ symmetric matrix, such that
\begin{align}
    K = L(L+I)^{-1} \Rightarrow L = K(I-K)^{-1},
\end{align}
whenever the left-hand side exists. 
In this case, letting $\mathbf X \sim \mathcal P$ be the random variable associated with the DPP, the probability of drawing any set $X$ is 
\begin{align}
    \mathbf P(\mathbf X = X) \propto \det L_X,
\end{align}
which in particular is not a marginal probability. Furthermore, the normalization constant is $Z(L) = \det L + I$. An interesting property is the following. Let $|\mathbf X|$ be the cardinality of $\mathbf X$. With $(\lambda_i)_{i=1,2,\ldots, m}$ the descending eigenvalues of $L$, $|\mathbf X|$ is distributed as the number of successes in $m$ Bernoulli trials with probability $\lambda_i/(\lambda_i + 1)$. In particular, it follows that $\mathbf E \left[ |\mathbf X | \right] = \sum_{i=1}^m \lambda_i/(\lambda_i + 1)$.\\ \\
In this example we will consider $m = 500$, and for each item $i$ we draw a uniformly distributed point $s_i$ in the unit square. To calculate $L$, we apply the standard squared exponential kernel $L(i,j) = \exp \{-\frac{1}{2} \Vert s_i - s_j \Vert^2 \}$. The expected number of points drawn from $\mathcal P_L$ is very close to 60 for any random selection of points under this model. We are interested in how many points are being drawn, and if a varied selection of points is being chosen, so similarly to the above example, our target statistic is the number of active points at any time, and the samplers are initiated at $S_0 = \{1,2,\ldots, m\},$ e.g. every point is active. In this case both the Tabu sampler and the Zanella mix well across the posterior, and the ESS/s of the log-energy are close, with a slight edge to the Tabu sampler, corresponding to an improvement in efficiency of about 50\%. However, when it comes to the generation of diverse point clouds, measured by the number of points chosen, the directed movement of the Tabu sampler leads to the sampler being around 6 times more efficient. In practice, the Zanella process is primarily modifying outliers with high impact, in probability, on the overall cloud, while the Tabu sampler uses its irreversible behaviour to more effectively change the composition. Neither sampler ever proposes more than 80 or fewer than 40 points, and the mean excursion length is 11.

%% file: Inputs/D_Numerical_Examples/HighOrder.tex
In the following sections we consider two examples of spaces where the order of the generators are infinite and finite, respectively. We compare our proposed dZZ and dCS with the Zanella process.

%% file: Inputs/D_Numerical_Examples/LatticeGaussian.tex
Extensions of the Gaussian distribution to discrete observations and spaces have received some attention, in particular via Boltzmann machines and lattice Gaussians.
The latter distribution has since the seminal paper \cite{gentry2008trapdoors} been applied within cryptography schemes, in particular because these lattice-based encryption methods appear resilient to brute-force algorithms given even significant quantum computing power (see \cite{follath2014gaussian} for extended discussion.) The hardness part of lattice-based encryption schemes is based on the difficulty of finding the shortest possible vector (the 'SVP' in encryption theory) given a lattice \cite{howe2016practical}. Furthermore, sampling these distributions require at least 100 bit floating point precision, adding additional constraints to the implementation of the sampling algorithms. Outside of quantum cryptography, the lattice Gaussian distribution has also seen use in coding theory \cite{ling2014achieving} and spatial econometrics \cite{anselin2001spatial}. Here we consider a simple toy case to illustrate the dynamics of the discrete Zig-Zag process and discrete coordinate sampler, and the Zanella process on spaces with higher-order generators.
Let $d > 1$ be a fixed integer, and consider a set of basis vectors $(\mathbf v_i)_{1 \leq i \leq d}$ in $\mathbb R^d$, which in general will be neither normalized nor mutually orthogonal. We then define a lattice $\mathbf B$ as the countably-infinite set of vectors
\begin{align}
    \mathbf B = \left \{ \sum_{i=1}^d \mathbf v_i z_i : \mathbf z = ( z_1, \cdots, z_d ) \in \mathbb Z^d \right \}
\end{align}
For simplicity, we denote by $B\mathbf z \in \mathbf B$ the point on the lattice given by the input vector $\mathbf z$. Given an arbitrary lattice $\mathbf B$, the centered lattice Gaussian has probability proportional to
\begin{align}
    \pi(\mathbf z) \propto \exp \left( -  \frac{\pi \Vert B\mathbf z \Vert_2^2 }{s^2} \right)
\end{align}
for some Gaussian parameter $s > 0$. For the distribution to be useful in cryptography, the variance parameter has to be moderately large; we take $s = 500$ in order to satisfy the requirements posited in the BLISS scheme of \cite{ducas2013lattice}. While the choice of lattice basis is fundamental in cryptographic applications, the extra cost of working on $\mathbf B$ is negligible. As such, we focus our attention on working with the natural basis of $\mathbb Z^d$. 

To compare the methods, we initially let $d = 3$. In this case, it is clear from the algorithmic constructions that the runtime of the Zanella process will be twice that of the dZZ, while the dZZ again will at least be thrice as slow as the dCS. To elucidate the difference, we first run the samplers until the algorithm time $T = 100,000$ and thin at intervals of $t = 1$, and subsequently run until the Zanella process and dCS have run for an equivalent amount of wall-clock time compared to the Zanella process (We re-run the experiments multiple times to verify the consistency in the comparison.) Our initialization point is $z_0 = (1000, 1000, 1000)$, which is far out in the tails of the target.

\begin{figure}
    \centering
    \includegraphics[width=\textwidth]{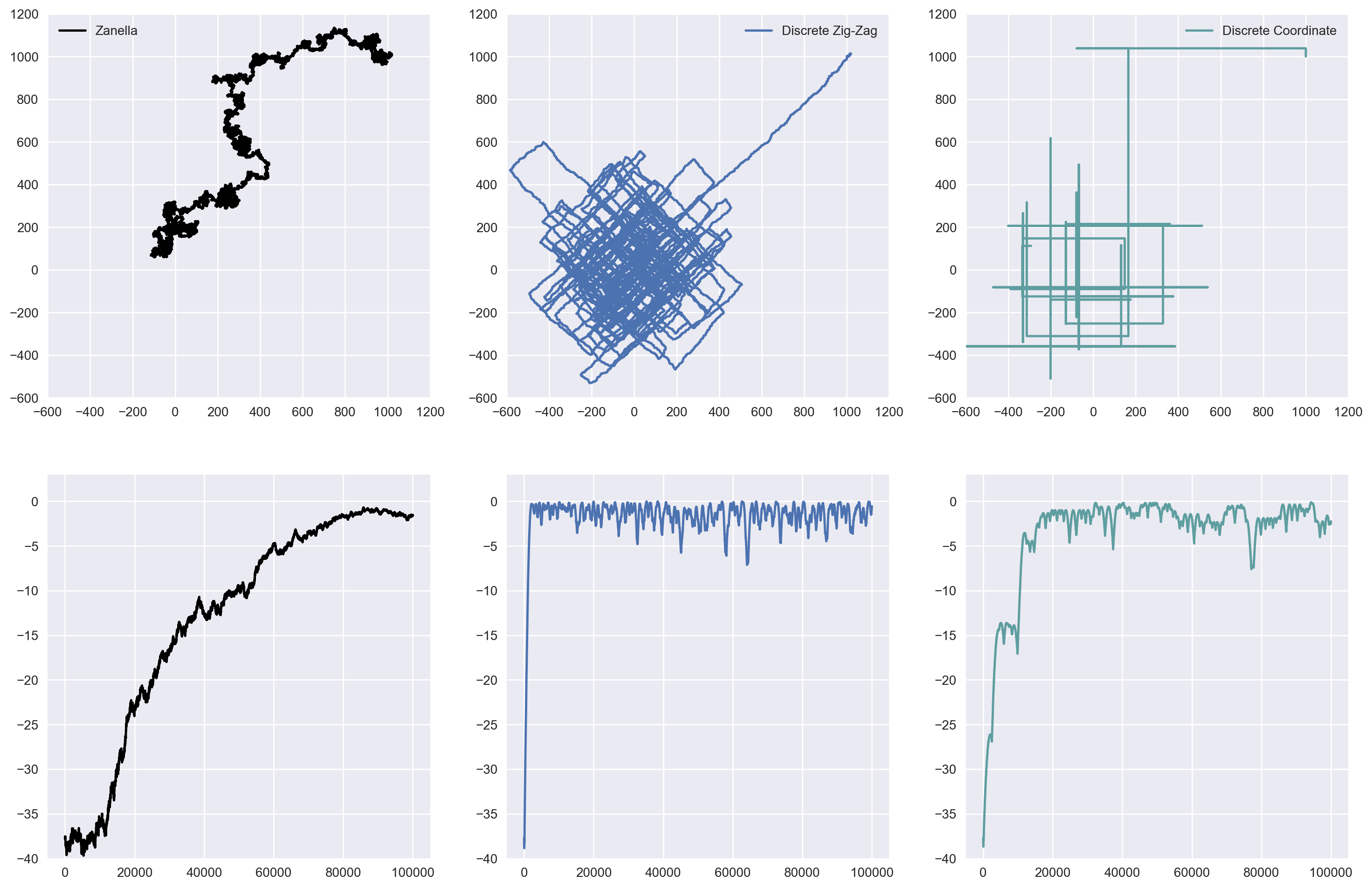}
    \caption{Sampling until internal time $T = 100,000.$ Upper row: Position of the first and second coordinate of the 3-dimensional lattice Gaussian. Lower row: traceplot of the log-probability for the distribution. }
    \label{fig:dpdmp_internal}
\end{figure}

\begin{figure}
    \centering
    \includegraphics[width=\textwidth]{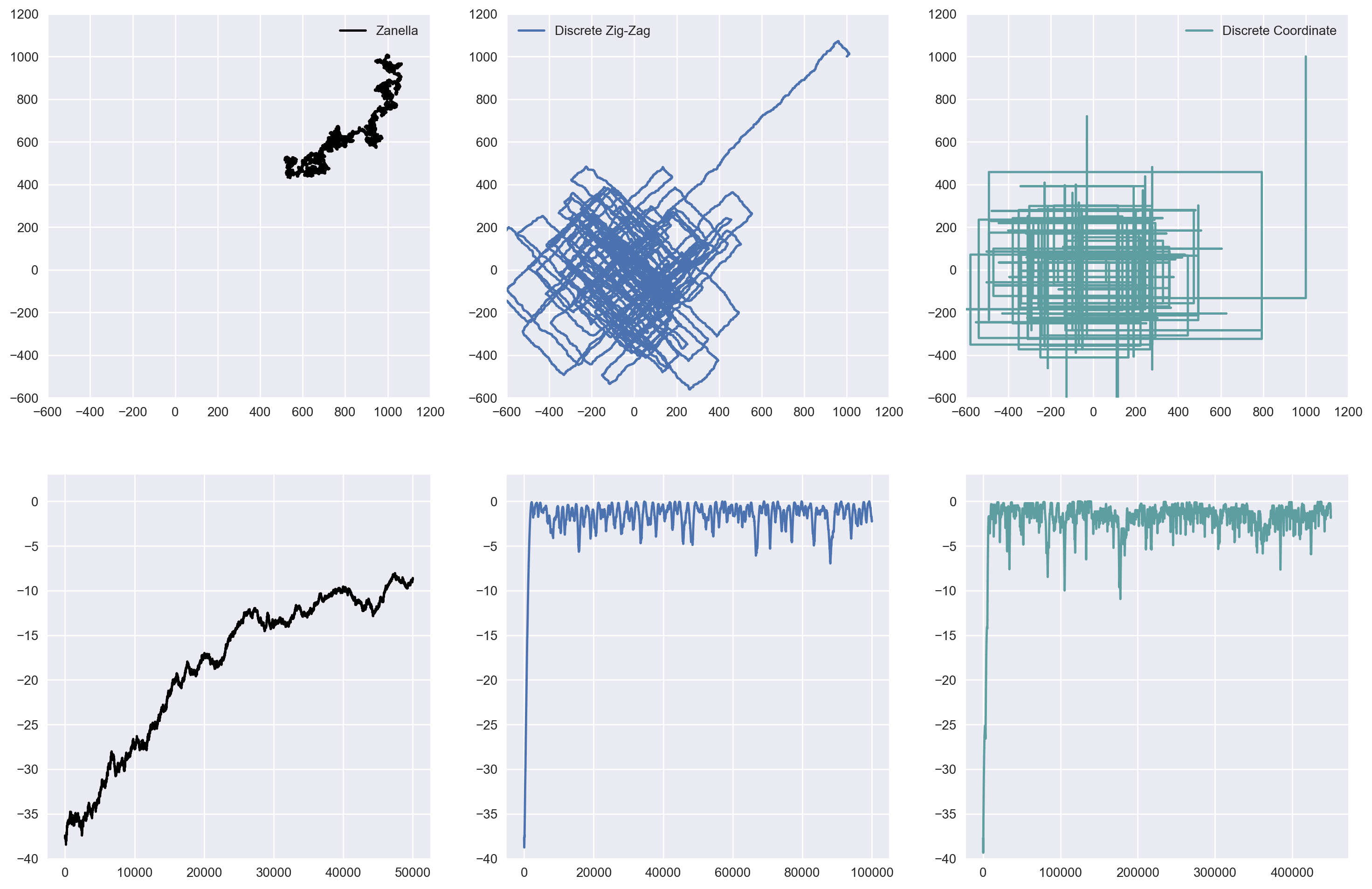}
    \caption{Sampling until walk-clock time is equivalent to the runtime to generate 100,000 samples from the discrete zig-zag sampler. Upper row: Position of the first and second coordinate of the 3-dimensional lattice Gaussian. Lower row: traceplot of the log-probability for the distribution. }
    \label{fig:dpdmp_external}
\end{figure}

The results are shown in Figure \ref{fig:dpdmp_internal} and \ref{fig:dpdmp_external}. The random-walk like structure of the Zanella process highlights the difficulty in designing good proposal mechanisms for the lattice Gaussian, since at any given point the change in probability at each neighbour is miniscule. The consequence is that the locally-balanced weighting function effectively loses its ability to distinguish meaningfully between neighbouring states, and the sampler begins to resemble a random walk Metropolis-Hastings sampler. We suspect that for heavy-tailed target distributions, the Zanella process will run into this problem more generally. 

In contrast to the Zanella process, the irreversible dZZ and dCS mix quickly, as their persistent behaviour drags them towards the mode of the distribution. For this particular target, a naive implementation (in the sense of not exploiting known modes, symmetries etc. of the target) of the dCS runs 4.5 times quicker than the dZZ, however the ESS/s is of the same order. 

%% file: Inputs/D_Numerical_Examples/LatticeGauge.tex
In particle physics, \textit{lattice gauge theory} (LGT) is a model of quantum field theory in which space is discretised onto a lattice. Calculating quantities of interest in lattice gauge theory involves evaluating high-dimensional integrals, and as such, Monte Carlo methods have been widely used in this area; see \cite{rebbi1983lattice} for an introduction. In the standard formulation, the state of an LGT model is a collection of `gauge field' variables, indexed by the edges of the lattice, i.e. $x = \{ x_{i, j} \}_{ (i, j) \in E ( L ) }$, where the gauge fields $x_{i, j}$ take values in some Lie group, e.g. the group of two-dimensional rotations, $SO ( 2 ) \cong \mathbb{S}^1$. It is occasionally of interest to go one step further, and also discretise \textit{the group}; this gives rise to a \textit{discrete lattice gauge theory}; see e.g. \cite{romers2007discrete}. We will apply our algorithms to a specific instance of a discrete LGT.

In particular, we focus on the setting where the lattice is a subset of $\mathbb{Z}^2$, the continuous group is $SO ( 2)$, and the discrete approximation is given by $\mathbb{Z}_p$, i.e. the integers $\mod p$. Our target distribution is defined as follows: for each $1 \times 1$ square $P$ in the lattice, with vertices $(i, j, k, l)$, define 
\begin{align}
    V ( P ) = 1 - \cos \left( \frac{ 2 \pi }{ p } \left[ x_{i, j} + x_{j, k} + x_{k, l} + x_{l, i} \right] \right).
\end{align}
For $\beta > 0$, we then define the target distribution $\pi$ as
\begin{align}
    \pi ( x ) \propto \exp \left( - \beta \sum_{P \in L} V ( P ) \right).
\end{align}
Our state space is then $\mathbb{Z}_P^{E ( L )}$, with generators given by `increase $x_{i,j}$ by $1$', for each edge $(i, j)$. For large $p$, these generators have high order, and so our PDMP-type algorithms are appropriate.

\begin{figure}[h]
    \centering
    \includegraphics[width=\textwidth]{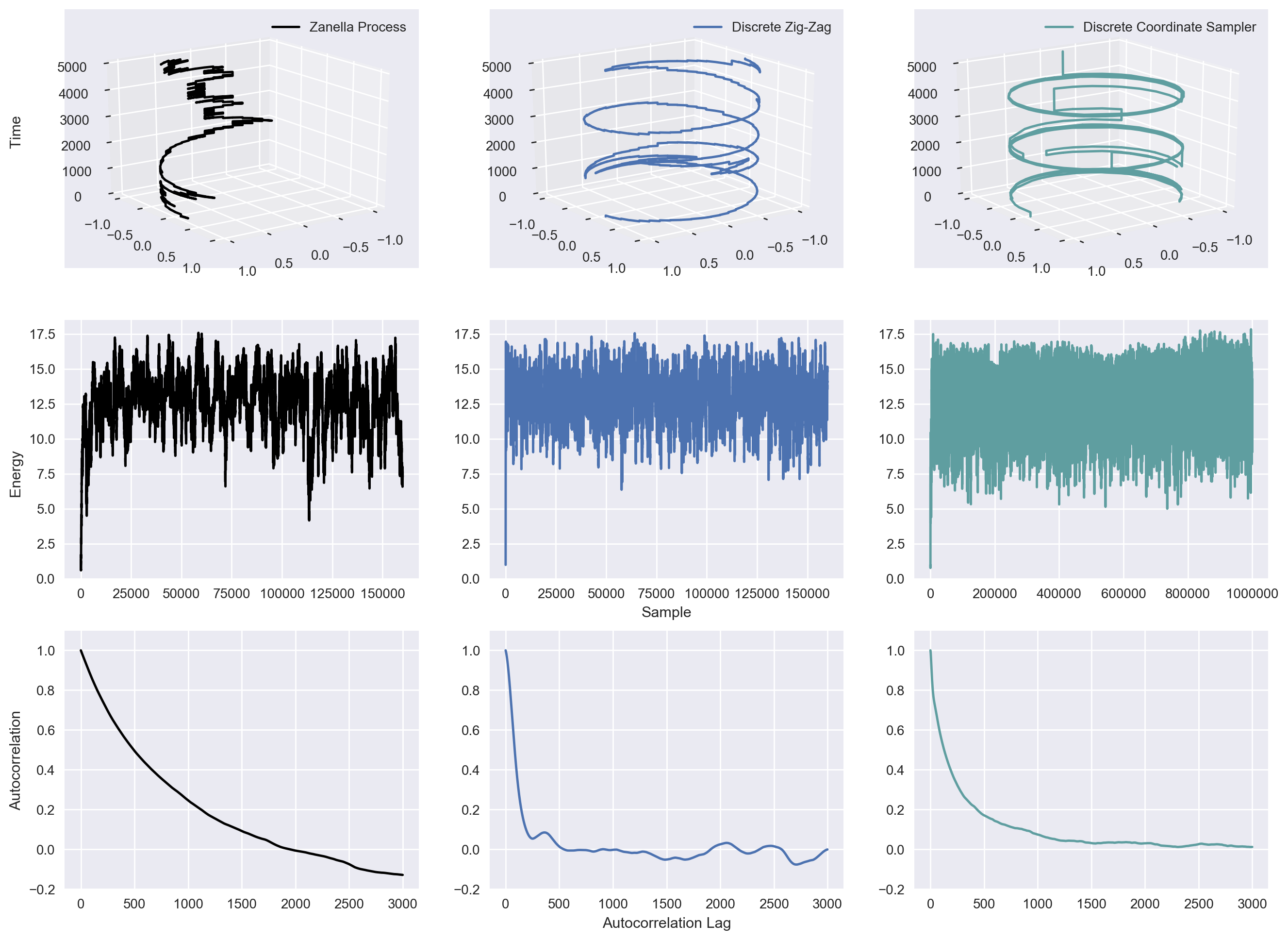}
    \caption{Sampling until wall-clock time corresponds to $10^6$ samples from the dCS. Upper row: Position on the circle of the first coordinate up until $t = 5,000$. Middle row: traceplot of the log-probability of the distribution. Lower row: Autocorrelation function up to order $3,000$, calculated after discarding the first 20\% of samples generated.}
    \label{fig:dpdmp_gauge}
\end{figure}

For comparison, we let $\beta = 1$, $p = 53$ and consider a $4 \times 4$ subset of $\mathbb{Z}^2$. This gives us $P = 9$ squares and 24 vertices, for a total of $53^{24}$ possible sample combinations, illustrating the challenges the discrete lattice gauge model poses even in very low dimensions. We ran each sampler for an equivalent amount of wall-clock time (relative to generating $10^6$ samples with the dCS), thinning approximately at each event of the sampler. Because of the dependencies across factors, the Zanella process and dZZ in this case generate an equal amount of samples at the same runtime, while we note that the Zanella process is thinning at twice the rate. To compare the resulting dynamics, we consider the following map of time and the first coordinate
\begin{align}
    (t, x_1) \mapsto \left (t, \cos \left (\frac{2 \pi}{p}(x_1 \mod{p}) \right ), \sin\left (\frac{2 \pi}{p}(x_1 \mod{p} \right )\right ),
\end{align}
and plot the resulting movement on the circle in the upper row of Figure \ref{fig:dpdmp_gauge}. The dynamics are very similar to the ones observed for the lattice Gaussian, with the Zanella process experiencing difficulty in circumnavigating the whole circle, but the overall energy still mixes reasonably well. In comparison, the dZZ showcases good persistent behaviour at all times as it applies the same generators recurrently, while the dCS interestingly enough occasionally covers the entire circle more than once whenever the coordinate is active. Comparing our new schemes, the dZZ decorrelates much quicker than the dCS, but also requires access to all states accessible via a generator at each iteration. The resulting ESS/s is therefore still slightly higher for the dCS compared to the dZZ, with approximately 13 and 10 times improvement on the Zanella process, respectively, on average. We also note that the advantage of the dCS relative to the dZZ only increases as the size of the lattice grows. 

%% file: outlook.tex
In this work, we have presented a collection of algorithms which can be used for continuous-time Monte Carlo sampling on structured, discrete spaces. We have established correctness of these algorithms, and demonstrated their performance on a collection of examples. A particular benefit of these samplers is that they each directly use the structure of the underlying discrete space, and do not rely upon relaxations or embeddings into continuous spaces to achieve their improved performance.

Going forward, we highlight three directions in which this line of work could be advanced. 

An important starting point would be to carry out a theoretical study of the quantitative convergence properties of these new samplers, and under which conditions we can expect them to mix quickly. On continuous state spaces, one generally believes that log-concavity is sufficient for most sensible algorithms to mix quickly; on discrete spaces, it is not yet clear whether there is a similarly easily-verified condition which could ensure fast mixing of `sensible' Markov chains on general state spaces (though see \cite{johnson2017discrete} for some quite specific steps in this direction). On a related note, there is a wealth of existing work which makes use of group theory and representation theory to study the mixing behaviour of Markov chains, typically to uniform distributions. It would be exciting to see whether these techniques could provide useful answers in the setting of sampling from non-uniform distributions over structured spaces.



At a directly practical level, there is work to be done on how these algorithms can fit into the broader Monte Carlo toolbox. For example, we have used a naive thinning scheme to calculate our sample algorithms; one can imagine that a Rao-Blackwellisation procedure along the lines of \cite{casella1996rao, douc2011vanilla} could provide some cheap variance reduction. One could also try to identify natural couplings for these processes, which could enable their use in the unbiased estimation schemes of \cite{jacob2017unbiased}. There is also ample opportunity to study how parallelism, adaptation schemes, and convergence diagnostics could play a role in further improving the performance of these algorithms.

Finally, we are excited to see how these techniques could give rise to improved algorithms for target distributions with a combination of discrete and continuous components. This could include statistical models with this flavour (e.g. mixture models, as in \cite{richardson1997bayesian, stephens2000bayesian, cappe2003reversible}), or artificially-discrete sampling methods like Parallel Tempering (e.g. \cite{earl2005parallel}). Recent work has shown that continuous-time sampling from continuous distributions can be done tractably and efficiently; it would be rewarding to connect those techniques with the methods in this paper to devise fully general-purpose samplers, in this regard.

%% file: Inputs/E_Appendix_Bibliography/InvarianceProofs.tex
\subsection{Proof Techniques}

In this section, consider a Markov Jump Process on a discrete state space $\Uc$, with jump rates $\lambda (u \to v)$, and generator given by
\begin{align}
    ( \Lc f ) ( u ) = \sum_{v \in \partial u} \lambda ( u \to v ) \left[ f ( v ) - f ( u ) \right].
\end{align}

For such processes, there are two standard routes to establishing invariance with respect to a given measure $\Pi$. 

\subsubsection{Detailed Balance}

The first of these is \textit{detailed balance}. An MJP is said to be in detailed balance with respect to $\Pi$ if
\begin{align}
    \forall f, g \in L^2 ( \Pi ), \quad \Eb_\Pi [ ( \Lc f ) ( u ) g ( u )] = \Eb_\Pi [ f ( u ) ( \Lc g) ( u )].
\end{align}
It is standard to show that this condition is sufficient for $\Pi$ to be an invariant measure for the MJP. By considering $f ( x ) = \mathbf{I} [ x = u ], g ( x ) = \mathbf{I} [ x = v ]$, one can show that it is necessary and sufficient that,
\begin{align}
    \forall u, v \in \Uc, \quad \Pi ( u ) \lambda ( u \to v ) = \Pi ( v ) \lambda ( v \to u ),
\end{align}
which can typically be verified by inspection.

\subsubsection{Skew-Detailed Balance}

The second such route is known as \textit{skew-detailed balance}. In this case, the space $\Uc$ is equipped with a map $S: \Uc \to \Uc$, satisfying $S ( S ( u ) ) = u $ for all $u$. Defining the operator $\Qc$ by
\begin{align}
    Q f ( u ) = f ( S ( u ) ),
\end{align}
the MJP is said to be in skew-detailed balance with respect to $( \Pi, \Qc )$ if
\begin{align}
    \forall f, g, \quad \Eb_\Pi [ ( \Qc \Lc f ) ( u ) g ( u )] = \Eb_\Pi [ f ( u ) ( \Qc \Lc g) ( u )].
\end{align}
In what follows, we derive some assumptions under which skew-detailed balance will hold. In particular, we require following `local conditions'
\begin{align}
    \forall u \in \Uc, \quad \Pi ( S ( u ) ) &= \Pi ( u ), \\
    \forall u, v \in \Uc, \quad  \Pi ( u ) \lambda ( u \to v ) &= \Pi ( S ( v ) ) \lambda ( S ( v ) \to S ( u ) ).
\end{align}
Moreover, defining
\begin{align}
    \Lambda ( u ) &= \sum_{v \in \partial u} \lambda ( u \to v ), 
\end{align}
we require the additional `semi-local condition'
\begin{align}
    \forall u \in \Uc, \quad \Lambda ( u ) = \Lambda ( S ( u ) ).
\end{align}

The local conditions are relatively standard, and crop up in e.g. the literature on non-reversible discrete-time Markov chains. However, the semi-local condition appears to be unique to the continuous-time setting, and may be of independent interest.

In any case, under these conditions, the MJP can be demonstrated to be skew-reversible, and thus leave $\Pi$ invariant.

We note that if an MJP satisfies the local conditions, but \textbf{not} the semi-local condition, it is possible to induce skew-detailed balance by including an extra jump type, as follows: at rate $\left[ \Lambda ( S ( u ) ) - \Lambda ( u ) \right]_+$, jump from $u$ to $S ( u )$. Indeed, this is how we derived the algorithms presented in this paper.

\clearpage

\subsection{Proofs for Individual Algorithms}

\subsubsection{Proof for the Zanella Process}

For the Zanella Process, one has that 
\begin{align}
    \pi ( x ) \lambda ( x \to y ) = \pi ( y ) \lambda ( y \to x ),
\end{align}
by construction, as the jump rates are defined using balancing functions. As such, by detailed balance, the algorithm leaves $\pi$ invariant.


\subsubsection{Proof for the SAW with Order-2 Generators}
For the SAW, we will show invariance by establishing that the process is skew-reversible with respect to the joint invariant measure $\Pi ( x, \alpha, \tau )$, and the involution $S ( x, \alpha, \tau ) = ( x, \alpha, - \tau )$. In particular, we claim that
\begin{align}
    \Pi ( x, \alpha, \tau ) \lambda ( ( x, \alpha, \tau ) \to ( \gamma \cdot x, \alpha^\gamma, \tau ) = \Pi ( \gamma \cdot x, \alpha^\gamma, - \tau ) \lambda ( ( \gamma \cdot x, \alpha^\gamma, - \tau ) \to ( x, \alpha, - \tau ) ).
\end{align}
As this transition leaves the probability mass functions of $\alpha$ and $\tau$ unchanged, we need only check that
\begin{align}
    \pi ( x ) \lambda ( ( x, \alpha, \tau ) \to ( \gamma \cdot x, \alpha^\gamma, \tau ) = \pi ( \gamma \cdot x) \lambda ( ( \gamma \cdot x, \alpha^\gamma, - \tau ) \to ( x, \alpha, - \tau ) ).
\end{align}
Expanding the jump probabilities, this is equivalent to 
\begin{align}
    \pi ( x ) g \left( \frac{ \pi ( \gamma \cdot x) }{ \pi ( x )} \right) \cdot \mathbf{I} [ \alpha ( \gamma ) = \tau ] = \pi ( \gamma \cdot x) g \left( \frac{ \pi (x) }{ \pi ( \gamma \cdot x )} \right) \cdot \mathbf{I} [ \alpha^\gamma ( \gamma ) = - \tau ].
\end{align}
The indicator functions in this expression are equal by construction, i.e. the move from $( x, \alpha, \tau )$ to $( \gamma \cdot x, \alpha^\gamma, \tau )$ is allowed exactly when the move from $( \gamma \cdot x, \alpha^\gamma, - \tau )$ to $( x, \alpha, - \tau )$ is allowed. Moreover, because the jump rates are given in terms of balancing functions, one has that
\begin{align}
    \pi ( x ) g \left( \frac{ \pi ( \gamma \cdot x) }{ \pi ( x )} \right) = \pi ( \gamma \cdot x) g \left( \frac{ \pi (x) }{ \pi ( \gamma \cdot x )} \right).
\end{align}
As such, the two sides are genuinely equal. The additional jump rates for transitioning between $( x, \alpha, \tau )$ and $S ( x, \alpha, \tau ) = ( x, \alpha, - \tau )$ ensure that $\Lambda ( ( x, \alpha, \tau ) ) = \Lambda ( S ( x, \alpha, \tau ) )$, and thus by skew-reversibility, the process leaves $\Pi$ invariant.

\subsubsection{Proof for the discrete Zig-Zag Process}
For the discrete Zig-Zag process, we will show invariance by first establishing that when only one generator is available, the process leaves $\Pi ( x, \theta)$ invariant. As the full discrete Zig-Zag process can be viewed as a superposition of the single-generator processes, the infinitesimal generator of the full process is simply the sum of the infinitesimal generators of the single-generator processes, and one can thus deduce that the full process leaves $\Pi$ invariant.

To estabish that the single-generator process leaves $\Pi$ invariant, we will show that the process is skew-reversible with respect to the joint invariant measure $\Pi ( x, \theta ( \gamma ) )$ and the involution $S ( x, \theta ( \gamma ) ) = ( x, -\theta ( \gamma ) )$. This claim is equivalent to the equality
\begin{align}
    \Pi ( x, \theta ) \cdot \lambda ( ( x, \theta ) \to ( y , \theta ) ) = \Pi ( y, - \theta ) \cdot \lambda ( (y, - \theta ) \to ( x, - \theta ) )
\end{align}
where $y = \gamma^{ \theta ( \gamma ) } \cdot x, \theta = \theta ( \gamma )$. As in the case of the SAW, the transition leaves the probability mass function of $\theta$ unchanged, and so we need only check that 
\begin{align}
    \pi ( x ) \cdot g \left( \frac{ \pi ( y ) }{ \pi ( x ) } \right) \cdot \mathbf{I} [ x \to y \text{ allowed by } \theta ]  = \pi ( y )  \cdot g \left( \frac{ \pi ( x ) }{ \pi ( y ) } \right) \cdot \mathbf{I} [ y \to x \text{ allowed by } - \theta ].
\end{align}
Again, the indicator functions are equal by construction, and the use of balancing functions guarantees that
\begin{align}
    \pi ( x ) \cdot g \left( \frac{ \pi ( y ) }{ \pi ( x ) } \right) = \pi ( y )  \cdot g \left( \frac{ \pi ( x ) }{ \pi ( y ) } \right).
\end{align}
As such, the two sides are genuinely equal. The additional jump rates for transitioning between $( x, \theta )$ and $S ( x, \theta ) = ( x, - \theta )$ ensure that $\Lambda ( ( x, \theta ) ) = \Lambda ( S ( x, \theta ) )$, and thus by skew-reversibility, the process leaves $\Pi$ invariant.

Now, let $\Lc^\gamma$ be the generator of the discrete Zig-Zag process which only uses the generator $\gamma$. The generator of the full process is then given by $\Lc = \sum_{\gamma \in \Gamma_0} \Lc^\gamma$. Since each of these generators is associated to a process which leaves $\Pi$ invariant, we see that for any $f \in L^2 ( \Pi )$, 
\begin{align}
    \Eb_\Pi \left[ ( \Lc^\gamma f ) ( x ) \right] &= 0 \quad \text{for all } \gamma \in \Gamma_0 \\
    \implies \sum_{ \gamma \in \Gamma_0 } \Eb_\Pi \left[ ( \Lc^\gamma f ) ( x ) \right] &= 0 \\
    \implies \Eb_\Pi \left[ \sum_{ \gamma \in \Gamma_0 } ( \Lc^\gamma f ) ( x ) \right] &= 0 \\
    \implies \Eb_\Pi \left[ ( \Lc f ) ( x ) \right] &= 0
\end{align}
and thus that the full process also leaves $\Pi$ invariant. 

Note that a simple adaptation of this proof allows for the design a correct variant of the discrete Zig-Zag process which prefers to use some generators more than others, by constructing a process with generator $\Lc_w = \sum_{ \gamma \in \Gamma_0} w ( \gamma ) \Lc^\gamma$ for some set of positive weights $w$. There may be scope for using such a process with an adaptive choice of $w$ to accelerate sampling. 

\subsubsection{Proof for the discrete Coordinate Sampler}

For the discrete Coordinate Sampler, we will show invariance by a direct computation, as the mechanism for changing the velocity is somewhat more elaborate than in the other algorithms. We begin by writing down the generator of the process:
\begin{align}
    (\Lc f ) ( x ) &= \delta ( x, v, \tau ) \cdot \left[ f ( v^\tau \cdot x, v, \tau ) - f ( x, v, \tau ) \right] \\
    &+ \rho ( x, v, \tau ) \cdot \sum_{w \in \Gamma} \frac{ \psi ( w ) \rho ( x, w, \tau ) }{ Z ( x ) } \left[ f ( x, w, -\tau ) - f ( x, v, \tau ) \right].
\end{align}
We begin by computing the expectation of the first part of the first term:
\begin{align}
    \Eb_\Pi \left[ \delta ( x, v, \tau ) \cdot f ( v^\tau \cdot x, v, \tau ) \right] &= \sum_{x, v, \tau} \pi ( x ) \psi ( v ) R ( \tau ) \delta ( x, v, \tau ) \cdot f ( v^\tau \cdot x, v, \tau ) \\
    &= \sum_{x, v, \tau} \pi ( x ) \psi ( v ) R ( \tau ) \delta ( x, v, \tau ) \cdot f ( v^\tau \cdot x, v, \tau ) \\
    &= \sum_{x, v, \tau} \pi ( x ) \psi ( v ) R ( \tau ) g \left( \frac{ \pi ( v^\tau \cdot x ) }{ \pi ( x )} \right) \cdot f ( v^\tau \cdot x, v, \tau ) \\
    &= \sum_{x, v, \tau} \pi ( v^\tau \cdot x ) \psi ( v ) R ( \tau ) g \left( \frac{ \pi ( x ) }{ \pi ( v^\tau \cdot x )} \right) \cdot f ( v^\tau \cdot x, v, \tau ) \\
    &= \sum_{y, v, \tau} \pi ( y ) \psi ( v ) R ( \tau ) g \left( \frac{ \pi ( v^{-\tau} \cdot y ) }{ \pi ( y )} \right) \cdot f ( y, v, \tau ) \\
    &= \sum_{y, v, \tau} \pi ( y ) \psi ( v ) R ( \tau ) \delta ( y, v, -\tau ) \cdot f ( y, v, \tau ) \\
    &= \Eb_\Pi \left[ \delta ( x, v, - \tau ) \cdot f ( x, v, \tau ) \right]
\end{align}
We can thus deduce that
\begin{align}
    \Eb_\Pi \left[ \delta ( x, v, \tau ) \cdot \left[ f ( v^\tau \cdot x, v, \tau ) - f ( x, v, \tau ) \right] \right] &= \Eb_\Pi \left[ \left\{ \delta ( x, v, - \tau ) - \delta ( x, v, \tau ) \right\} \cdot f ( x, v, \tau ) \right]
\end{align}
We now focus on computing the expectation of the first part of the second term:
\begin{align}
    &\Eb_\Pi \left[ \rho ( x, v, \tau ) \cdot \sum_{w \in \Gamma} \frac{ \psi ( w ) \rho ( x, w, \tau ) }{ Z ( x ) } f ( x, w, -\tau ) \right] \\
    &= \sum_{x, v, \tau, w} \pi ( x ) \psi ( v ) R ( \tau ) \cdot \rho ( x, v, \tau ) \cdot \frac{ \psi ( w ) \rho ( x, w, \tau ) }{ Z ( x ) } f ( x, w, -\tau ) \\
    &= \sum_{x, v, \sigma, w} \pi ( x ) \psi ( w ) R ( \sigma ) \cdot \rho ( x, w, - \sigma ) \cdot \frac{ \psi ( v ) \rho ( x, v, -\sigma ) }{ Z ( x ) } f ( x, v, \sigma ) \\
    &= \sum_{x, v, \sigma,} \pi ( x )  R ( \sigma ) \psi ( v ) \rho ( x, v, -\sigma ) f ( x, v, \sigma )  \sum_w \frac{ \psi ( w ) \rho ( x, w, - \sigma ) }{ Z ( x )} \\
    &= \sum_{x, v, \sigma,} \pi ( x )  R ( \sigma ) \psi ( v ) \rho ( x, v, -\sigma ) f ( x, v, \sigma ) \\
    &= \Eb_\Pi \left[ \rho ( x, v, - \tau ) f ( x, v, \tau ) \right]
\end{align}
and thus that the second term has expectation
\begin{align}
    &= \Eb_\Pi \left[ \left\{ \rho ( x, v, - \tau ) - \rho ( x, v, \tau ) \right\} f ( x, v, \tau ) \right].
\end{align}
Now, by noting that
\begin{align}
    \rho ( x, v, - \tau ) - \rho ( x, v, \tau ) &= \left[ \delta ( x, v, \tau ) - \delta ( x, v, -\tau ) \right]_+ - \left[ \delta ( x, v, - \tau ) - \delta ( x, v, \tau ) \right]_+ \\
    &= \delta ( x, v, \tau ) - \delta ( x, v, -\tau ),
\end{align}
we can add together the two expectation terms, and see that $\Eb_\Pi \left[ ( \Lc f ) ( x, v, \tau ) \right] = 0$. Thus, $\Pi$ is an invariant measure for the chain.    